\documentclass[fleqn]{llncs} 
\usepackage{hyperref}
\usepackage{amsmath} 
\usepackage{version} 
\usepackage{ipaa} 
\newcommand{\doi}[1]{{doi:\href{http://doi.org/#1}{\nolinkurl{#1}}}\rmFullStop}

\newcommand{\eprintlnk}[1]{{\href{#1}{Electronic version}}\rmFullStop}
\renewcommand{\url}[1]{{\href{#1}{\nolinkurl{#1}}}\rmFullStop}
\newcommand*{\rmFullStop}{\rmifnextchar{.}{}{}}
\makeatletter
\newcommand{\rmifnextchar}[3]{%
  \begingroup
  \ltx@LocToksA{\endgroup#2}%
  \ltx@LocToksB{\endgroup#3}%
  \ltx@ifnextchar{#1}{%
    \def\next{\the\ltx@LocToksA}%
    \afterassignment\next
    \let\scratch= %
  }{%
    \the\ltx@LocToksB
  }%
}
\makeatother

\title{Imperative Process Algebra with Abstraction}
\author{C.A. Middelburg}
\institute{Informatics Institute, Faculty of Science, University of
           Amsterdam \\
           Science Park~904, 1098~XH Amsterdam, the Netherlands \\
           \email{C.A.Middelburg@uva.nl}}
  
\begin{document}
\maketitle

\begin{abstract}
% 119 %
This paper introduces an imperative process algebra based on ACP 
(Algebra of Communicating Processes). 
Like other imperative process algebras, this process algebra deals with 
processes of the kind that arises from the execution of imperative 
programs.
It distinguishes itself from already existing imperative process 
algebras among other things by supporting abstraction from actions that 
are considered not to be visible. 
The support of abstraction of this kind opens interesting application 
possibilities of the process algebra.
This paper goes briefly into the possibility of information-flow 
security analysis of the kind that is concerned with the leakage of 
confidential data.
For the presented axiomatization, soundness and semi-completeness 
results with respect to a notion of branching bisimulation equivalence 
are established.
\begin{classcode}
D.2.4, D.4.6, F.1.1, F.1.2, F.3.1. 
\end{classcode}
\end{abstract}

\section{Introduction}
\label{sect-intro}

Generally speaking, process algebras focus on the main role of a 
reactive system, namely maintaining a certain ongoing interaction with 
its environment.
Reactive systems contrast with transformational systems.
A transformational system is a system whose main role is to produce, 
without interruption by its environment, output data from input data.%
\footnote
{The terms reactive system and transformational system are used here
 with the meaning given in~\cite{HP85a}.}
In general, early computer-based systems were transformational systems.
Nowadays, many systems are composed of both reactive components and 
transformational components. 
A process carried out by such a system is a process in which data are 
involved.
Usually, the data change in the course of the process, the process 
proceeds at certain stages in a way that depends on the changing data, 
and the interaction of the process with other processes consists of 
communication of data.

This paper introduces an extension of ACP~\cite{BK84b} with features 
that are relevant to processes of the kind referred to above.
The extension concerned is called \deACPet. 
Its additional features include assignment actions to change the data in 
the course of a process, guarded commands to proceed at certain stages 
of a process in a way that depends on the changing data, and data
parameterized actions to communicate data between processes. 
The processes of the kind that \deACPet\ is concerned with are 
reminiscent of the processes that arise from the execution of imperative 
programs.
In~\cite{NP97a}, the term imperative process algebra was coined for 
process algebras like \deACPet.
Other imperative process algebras are VPLA~\cite{HI93a}, 
IPAL~\cite{NP97a}, CSP$_\sigma$~\cite{CH09a}, AWN~\cite{Feh12a}, and the
unnamed process algebra introduced in~\cite{BLSW20a}.

\deACPet\ distinguishes itself from those imperative process algebras by 
being the only one with the following three properties:
(1)~it supports abstraction from actions that are considered not to be 
visible;
(2)~a verification of the equivalence of two processes in its semantics 
is automatically valid in any semantics that is fully abstract with 
respect to some notion of observable behaviour (cf.~\cite{GW96a});
(3)~it offers the possibility of equational verification of process
equivalence.
CSP$_\sigma$ is the only one of the above-mentioned imperative process
algebras that has property~(1) and none of them has property~(2).
\deACPet\ is probably unique in being the only imperative process 
algebra with properties~(1), (2) and~(3).

Property~(1) is achieved by providing a special constant (called the 
silent step constant), special operators (called abstraction operators), 
and an appropriate notion of equivalence of processes in the semantics 
of \mbox{\deACPet}.
Property~(2) is achieved by using a notion of branching bisimulation
equivalence~\cite{GW96a} for the equivalence of processes in the 
semantics of \deACPet.
Property~(3) is achieved by providing an equational axiomatization of 
the equivalence concerned.

Property~(1) is essential for the verification of properties concerning 
the external behaviour of a system.
Property~(2) is desirable for such verifications in applications where
the final word on what exactly is observable behaviour has not been 
pronounced.
This means that \deACPet\ is an interesting process algebra for the 
verification of properties concerning the external behaviour of a system
whose description calls for an imperative process algebra.
It makes \deACPet, among other things, suitable for the verification of 
properties concerning the information-flow security of a system in which 
confidential and non-confidential data, contained in state components of 
the system, are looked up and changed and an ongoing interaction with 
the environment of the system is maintained.

A great part of the work done on information-flow security is concerned 
with secure information flow in programs, where information flow in a 
program is considered secure if information derivable from the 
confidential data contained in its high-security variables cannot be 
inferred from the non-confidential data contained in its low-security 
variables (see e.g.~\cite{VIS96a,SV98a,BC02a,NCC06a,Boh09a}).
A notable exception is the work done in a process-algebra setting, where 
the focus has shifted from programs to processes of the kind to which 
programs in execution belong and where the information flow in a process 
is usually considered secure if information derivable from confidential 
actions cannot be inferred from non-confidential actions
(see e.g.~\cite{FG95a,RS99a,Bos04a,Low04a}). 

Recent work done on information-flow security in a process-algebra 
setting occasionally deals with the data-oriented notion of secure 
information flow, but on such occasions program variables are always 
mimicked by processes (see e.g.~\cite{FRS05a,HY07a}). 
A suitable imperative process algebra would obviate the need to mimic
program variables.
This state of affairs motivated the development of \deACPet.
This paper also shows how \deACPet\ can be used for information-flow 
security analysis of the kind that is concerned with the leakage of 
confidential data.

The development of \deACPet\ was primarily aimed at obtaining an 
imperative process algebra with the properties that are designated above 
as essential and desirable for the verification of properties concerning 
the external behaviour of a system.
The starting point of the development of \deACPet\ is 
\ACPet~\cite[Section~5.3]{BW90}, which is a non-imperative process 
algebra with these properties.
This makes it a convenient starting point in view of the primary aim of 
the development.

\deACPet\ is closely related to \deACPei~\cite{BM19b}.
The main differences between them can be summarized as follows:
(a)~only the former supports abstraction from actions that are 
considered not to be visible and
(b)~only the latter has an iteration operator.
This paper introduces, in addition to \deACPet, guarded linear recursion 
in the setting of \deACPet.
The set of processes that can be defined by means of the operators of 
\deACPet\ extended with the iteration operator is a proper subset of the 
set of processes that can be defined by means of guarded linear 
recursion in the setting of \deACPet. 
Therefore, (a)~should be considered the important difference.
However, using the semantics of \deACPei\ as presented in~\cite{BM19b} as
the starting point of the semantics of \deACPet\ turned out to result in a
semantics that is too complicated to establish the soundness and 
semi-completeness results.

This paper is organized as follows.
First, a survey of the algebraic theory \ACPet, which is the extension 
of \ACP\ with the empty process constant $\ep$ and the silent step 
constant $\tau$, is given (Section~\ref{sect-ACPet}).
Next, the algebraic theory \deACPet\ is introduced as an extension of 
\ACPet\ (Section~\ref{sect-deACPet}) and guarded linear recursion in the 
setting of \deACPet\ is treated (Section~\ref{sect-deACPetr}).
After that, a structural operational semantics of \deACPet\ is presented 
and a notion of branching bisimulation equivalence based on this 
semantics is defined (Section~\ref{sect-semantics}).
Following this, the reasons for two relatively uncommon choices made in 
the preceding sections are clarified (Section~\ref{sect-interlude}).
Then, results concerning the soundness and (semi-)completeness of the 
given axiomatization with respect to branching bisimulation equivalence 
are established (Section~\ref{sect-sound-compl}).
Thereafter, it is explained how \deACPet\ can be used for 
information-flow security analysis of the kind that is concerned with 
the leakage of confidential data (Section~\ref{sect-info-flow}).
Finally, some concluding remarks are made 
(Section~\ref{sect-conclusions}).

There is also an appendix in which, for comparison, an alternative 
structural operational semantics of \deACPet\ is presented and a notion 
of branching bisimulation equivalence based on this alternative 
structural operational semantics is defined.
The alternative in question is the above-mentioned result of using the 
structural operational semantics of \deACPei\ as the starting point.

Section~\ref{sect-ACPet}, Section~\ref{sect-deACPet}, and the appendix 
mainly extend the material in Section~2, Section~3, and Section~4, 
respectively, of~\cite{BM19b}.
Portions of that material have been copied near verbatim or slightly 
modified.

\section{\ACP\ with Empty Process and Silent Step}
\label{sect-ACPet}

In this section, \ACPet\ is presented.
\ACPet\ is \ACP~\cite{BK84b} extended with the empty process constant 
$\ep$ and the silent step constant $\tau$ as 
in~\cite[Section~5.3]{BW90}.
In \ACPet, it is assumed that a fixed but arbitrary finite set $\Act$ of 
\emph{basic actions}, with $\tau,\dead,\ep \not\in \Act$, and a fixed 
but arbitrary commutative and associative \emph{communication} function 
$\funct{\commf}
 {(\Act \union \set{\tau,\dead}) \x (\Act \union \set{\tau,\dead})}
 {(\Act \union \set{\tau,\dead})}$, 
such that $\commf(\tau,a) = \dead$ and $\commf(\dead,a) = \dead$
for all $a \in \Act \union \set{\tau,\dead}$, have been given.
Basic actions are taken as atomic processes.
The function $\commf$ is regarded to give the result of synchronously
performing any two basic actions for which this is possible, and to be 
$\dead$ otherwise.
Henceforth, we write $\Actt$ for $\Act \union \set{\tau}$.

The algebraic theory \ACPet\ has one sort: the sort $\Proc$ of 
\emph{processes}.
This sort is made explicit to anticipate the need for many-sortedness 
later on. 
The algebraic theory \ACPet\ has the following constants and operators 
to build terms of sort~$\Proc$:
\begin{itemize}
\item
for each $a \in \Act$, the \emph{basic action} constant 
$\const{a}{\Proc}$;
\item
the \emph{silent step} constant $\const{\tau}{\Proc}$;
\item
the \emph{inaction} constant $\const{\dead}{\Proc}$;
\item
the \emph{empty process} constant $\const{\ep}{\Proc}$;
\item
the binary \emph{alternative composition} operator 
$\funct{\altc}{\Proc \x \Proc}{\Proc}$;
\item
the binary \emph{sequential composition} operator 
$\funct{\seqc}{\Proc \x \Proc}{\Proc}$;
\item
the binary \emph{parallel composition} operator 
$\funct{\parc}{\Proc \x \Proc}{\Proc}$;
\item
the binary \emph{left merge} operator 
$\funct{\leftm}{\Proc \x \Proc}{\Proc}$;
\item
the binary \emph{communication merge} operator 
$\funct{\commm}{\Proc \x \Proc}{\Proc}$;
\item
for each $H \subseteq \Act$ and for $H = \Actt$, 
the unary \emph{encapsulation} operator 
$\funct{\encap{H}}{\Proc}{\Proc}$;
\item
for each $I \subseteq \Act$, 
the unary \emph{abstraction} operator 
$\funct{\abstr{I}}{\Proc}{\Proc}$.
\end{itemize}
It is assumed that there is a countably infinite set $\cX$ of variables 
of sort $\Proc$, which contains $x$, $y$ and $z$.
Terms are built as usual.
Infix notation is used for the binary operators.
The following precedence conventions are used to reduce the need for
parentheses: the operator ${} \seqc {}$ binds stronger than all other 
binary operators and the operator ${} \altc {}$ binds weaker than all 
other binary operators.

The constants of \ACPet\ can be explained as follows ($a \in \Act$):
\begin{itemize}
\item
$a$ denotes the process that performs the observable action $a$ and 
after that terminates successfully;
\item
$\tau$ denotes the process that performs the unobservable action $\tau$ 
and after that terminates successfully;
\item
$\ep$ denotes the process that terminates successfully without 
performing any action;
\item
$\dead$ denotes the process that cannot do anything, it cannot even 
terminate successfully.
\end{itemize}
Let $t$ and $t'$ be closed \ACPet\ terms denoting processes $p$ and $p'$,
respectively, 
let $H \subseteq \Act$ or $H = \Actt$, and let $I \subseteq \Act$.
Then the operators of \ACPet\ can be explained as follows:
\begin{itemize}
\item
$t \altc t'$ denotes the process that behaves either as $p$ or as $p'$, 
where the choice between the two is resolved at the instant that one of 
them performs its first action or terminates successfully without 
performing any action, and not before;
\item
$t \seqc t'$\, denotes the process that first behaves as $p$ and 
following successful termination of $p$ behaves as $p'$;
\item
$t \parc t'$ denotes the process that behaves as $p$ and $p'$ in 
parallel, by which is meant that, each time an action is performed, 
either a next action of $p$ is performed or a next action of $p'$ is 
performed or a next action of $p$ and a next action of $p'$ are 
performed synchronously --- successful termination may take place
at any time that both $p$ and $p'$ can terminate successfully;
\item
$t \leftm t'$ denotes the same process as $t \parc t'$, except that it 
starts with performing an action of $p$;
\item
$t \commm t'$ denotes the same process as $t \parc t'$, except that it 
starts with performing an action of $p$ and an action of $p'$ 
synchronously;
\item
$\encap{H}(t)$ denotes the process that behaves the same as $p$, except 
that actions from $H$ are blocked from being performed;
\item
$\abstr{I}(t)$ denotes the process that behaves the same as $p$, except 
that actions from $I$ are turned into the unobservable action $\tau$.
\end{itemize}

The operators $\leftm$ and $\commm$ are of an auxiliary nature.
They make a finite axiomatization of \ACPet\ possible.

The operator $\encap{\Actt}$ can also be explained as 
follows:
$\encap{\Actt}\!(t)$ denotes the process that behaves the same as $\ep$ 
if $t$ denotes a process that has the option to behave the same as $\ep$ 
and it denotes the process that behaves the same as $\dead$ otherwise.
In~\cite[Section~5.3]{BW90}, the symbol $\surd$ is used 
instead of $\encap{\Actt}$.

% \pagebreak[2]
The axioms of \ACPet\ are presented in Table~\ref{axioms-ACPet}.
\begin{table}[!t]
\caption{Axioms of \ACPet}
\label{axioms-ACPet}
\begin{eqntbl}
\begin{axcol}
x \altc y = y \altc x                                & & \axiom{A1} \\
(x \altc y) \altc z = x \altc (y \altc z)            & & \axiom{A2} \\
x \altc x = x                                        & & \axiom{A3} \\
(x \altc y) \seqc z = x \seqc z \altc y \seqc z      & & \axiom{A4} \\
(x \seqc y) \seqc z = x \seqc (y \seqc z)            & & \axiom{A5} \\
x \altc \dead = x                                    & & \axiom{A6} \\
\dead \seqc x = \dead                                & & \axiom{A7} \\
x \seqc \ep = x                                      & & \axiom{A8} \\
\ep \seqc x = x                                      & & \axiom{A9} 
\eqnsep
x \parc y = x \leftm y \altc y \leftm x \altc x \commm y \altc
\encap{\Actt}\!(x) \seqc \encap{\Actt}\!(y)          & & \axiom{CM1E} \\
\ep \leftm x = \dead                                 & & \axiom{CM2E} \\
\alpha \seqc x \leftm y = \alpha \seqc (x \parc y)   & & \axiom{CM3}  \\
(x \altc y) \leftm z = x \leftm z \altc y \leftm z   & & \axiom{CM4}  \\
\ep \commm x = \dead                                 & & \axiom{CM5E} \\
x \commm \ep = \dead                                 & & \axiom{CM6E} \\
a \seqc x \commm b \seqc y = \commf(a,b) \seqc (x \parc y) 
                                                     & & \axiom{CM7}  \\
(x \altc y) \commm z = x \commm z \altc y \commm z   & & \axiom{CM8}  \\
x \commm (y \altc z) = x \commm y \altc x \commm z   & & \axiom{CM9}  
\eqnsep
\encap{H}(\ep) = \ep                                   & & \axiom{D0} \\
\encap{H}(\alpha) = \alpha      & \mif \alpha \notin H   & \axiom{D1} \\ 
\encap{H}(\alpha) = \dead       & \mif \alpha \in H      & \axiom{D2} \\
\encap{H}(x \altc y) = \encap{H}(x) \altc \encap{H}(y) & & \axiom{D3} \\
\encap{H}(x \seqc y) = \encap{H}(x) \seqc \encap{H}(y) & & \axiom{D4}
\eqnsep
\abstr{I}(\ep) = \ep                                   & & \axiom{T0} \\
\abstr{I}(\alpha) = \alpha      & \mif \alpha \notin I   & \axiom{T1} \\
\abstr{I}(\alpha) = \tau        & \mif \alpha \in I      & \axiom{T2} \\
\abstr{I}(x \altc y) = \abstr{I}(x) \altc \abstr{I}(y) & & \axiom{T3} \\
\abstr{I}(x \seqc y) = \abstr{I}(x) \seqc \abstr{I}(y) & & \axiom{T4} 
\eqnsep                                                                 
\alpha \seqc (\tau \seqc (x \altc y) \altc x) = \alpha \seqc (x \altc y)
                                                       & & \axiom{BE} 
\end{axcol}
\end{eqntbl}
\end{table}
In these equations, $a$, $b$, and $\alpha$ stand for arbitrary constants 
of \ACPet\ other than $\ep$,\, $H$ stands for an arbitrary subset of 
$\Act$ or the set $\Actt$, and $I$ stands for an arbitrary subset
\linebreak[2] of $\Act$.
So, CM3, CM7, D0--D4, T0--T4, and BE are actually axiom schemas.
In this paper, axiom schemas will usually be referred to as axioms.

The occurrence of the strange-looking term 
$\encap{\Actt}\!(x) \seqc \encap{\Actt}\!(y)$ in axiom CM1E deserves 
some explanation.
This term is needed to handle successful termination in the presence of 
$\ep$:
it stands for the process that behaves the same as $\ep$ if both $x$ and 
$y$ stand for a process that has the option to behave the same as $\ep$ 
and it stands for the process that behaves the same as $\dead$ 
otherwise.

Notice that there are no operators $\encap{H}$ for $H \subset \Actt$ 
with $\tau \in H$ in \ACPet.
If one or more of them were present, the equation 
$\alpha \seqc \dead = \alpha$ 
would be derivable from the axioms of \ACPet.

In the sequel, the notation $\Altc{i=1}{n} t_i$, where $n \geq 1$, will 
be used for right-nested alternative compositions.
For each $n \in \Natpos$,%
\footnote
{We write $\Natpos$ for the set $\set{n \in \Nat \where n \geq 1}$ of 
 positive natural numbers.} 
the term $\Altc{i = 1}{n} t_i$ is defined by induction on $n$ as 
follows:
$\Altc{i = 1}{1} t_i = t_1$ and 
$\Altc{i = 1}{n + 1} t_i = t_1 \altc \Altc{i = 1}{n} t_{i+1}$.
In addition, the convention will be used that 
$\Altc{i = 1}{0} t_i = \dead$.

\section{Imperative \ACPet}
\label{sect-deACPet}

In this section, \deACPet, imperative \ACPet, is presented.
This extension of \ACPet\ has been inspired by~\cite{BM09d}.
It extends \ACPet\ with features that are relevant to processes in which
data are involved, such as guarded commands (to deal with processes that 
only take place if some data-dependent condition holds), data 
parameterized actions (to deal with process interactions with data 
transfer), and assignment actions (to deal with data that change in the 
course of a process).

In \deACPet, it is assumed that the following has been given with 
respect to data:
\begin{itemize}
\item
a many-sorted signature $\sign_\gD$ that includes:
\begin{itemize}
\item
a sort $\Data$ of \emph{data} and
a sort $\Bool$ of \emph{booleans};
\item
constants of sort $\Data$ and/or operators with result sort $\Data$;
\item
constants $\Btrue$ and $\Bfalse$ of sort $\Bool$ and
operators with result sort $\Bool$;
\end{itemize}
\item
a minimal algebra $\gD$ of the signature $\sign_\gD$ in which 
the carrier of sort $\Bool$ has cardinality $2$ and 
the equation $\Btrue = \Bfalse$ does not hold.
\end{itemize}
We write $\DataVal$ for the set of all closed terms over the signature
$\sign_\gD$ that are of sort~$\Data$.
The sort $\Bool$ is assumed to be given in order to make it possible for 
operators to serve as predicates.

It is also assumed that a finite or countably infinite set $\ProgVar$ of 
\emph{flexible variables} has been given.
A flexible variable is a variable whose value may change in the course 
of a process.%
\footnote
{The term flexible variable is used for this kind of variables in 
 e.g.~\cite{Sch97a,Lam94a}.} 
Typical examples of flexible variables are the program variables known 
from imperative programming.
An \emph{evaluation map} is a function from $\ProgVar$ to $\DataVal$. 
We write $\EvalMap$ for the set of all evaluation maps.

The algebraic theory \deACPet\ has the following sorts: 
the sort $\Proc$ of \emph{processes},
the sort $\Cond$ of \emph{conditions}, and
the sorts from $\sign_\gD$.

It is assumed that there are countably infinite sets of variables of 
sort $\Cond$ and $\Data$ and that the sets of variables of sort $\Proc$, 
$\Cond$, and $\Data$ are mutually disjoint and disjoint from $\ProgVar$.

Below, the constants and operators of \deACPet\ are introduced.
The operators of \deACPet\ include two variable-binding operators.
The formation rules for \deACPet\ terms are the usual ones for the 
many-sorted case (see e.g.~\cite{ST99a,Wir90a}) and in addition the 
following rule:
\begin{itemize}
\item
if $O$ is a variable-binding operator 
$\funct{O}{S_1 \x \ldots \x S_n}{S}$ that binds a variable of sort~$S'$,
$t_1,\ldots,t_n$~are terms of sorts $S_1,\ldots,S_n$, respectively, and 
$X$ is a variable of sort $S'$, then $O X (t_1,\ldots,t_n)$ is a term of 
sort $S$.
\end{itemize}
An extensive formal treatment of the phenomenon of variable-binding 
operators can be found in~\cite{PS95a}.

\deACPet\ has the constants and operators from $\sign_\gD$ to build 
terms of the sorts from $\sign_\gD$ --- which include the sort $\Bool$ 
and the sort $\Data$ --- and in addition the following constants to 
build terms of sort $\Data$:
\begin{itemize}
\item
for each $v \in \ProgVar$, the \emph{flexible variable} constant 
$\const{v}{\Data}$.
\end{itemize}
We write $\DataTerm$ for the set of all closed \deACPet\ terms of sort 
$\Data$.

Evaluation maps are intended to provide the data values assigned to 
flexible variables when an \deACPet\ term of sort $\Data$ is evaluated.
However, in order to fit better in an algebraic setting, they provide 
closed terms over the signature $\sign_\gD$ that denote those data 
values instead.
The requirement that $\gD$ is a minimal algebra guarantees that each 
data value can be represented by a closed term. 

\deACPet\ has the following constants and operators to build terms of 
sort~$\Cond$:
\begin{itemize}
\item
the binary \emph{equality} operator
$\funct{\Leq}{\Bool \x \Bool}{\Cond}$;
\item
the binary \emph{equality} operator
$\funct{\Leq}{\Data \x \Data}{\Cond}$;%
\footnote
{The overloading of $=$ can be trivially resolved if $\sign_\gD$ is
 without overloaded symbols.}
\item
the \emph{truth} constant $\const{\True}{\Cond}$;
\item
the \emph{falsity} constant $\const{\False}{\Cond}$;
\item
the unary \emph{negation} operator $\funct{\Lnot}{\Cond}{\Cond}$;
\item
the binary \emph{conjunction} operator 
$\funct{\Land}{\Cond \x \Cond}{\Cond}$;
\item
the binary \emph{disjunction} operator 
$\funct{\Lor}{\Cond \x \Cond}{\Cond}$;
\item
the binary \emph{implication} operator 
$\funct{\Limpl}{\Cond \x \Cond}{\Cond}$;
\item
the unary variable-binding \emph{universal quantification} operator 
$\funct{\forall}{\Cond}{\Cond}$ that binds a variable of sort $\Data$; 
\item
the unary variable-binding \emph{existential quantification} operator 
$\funct{\exists}{\Cond}{\Cond}$ that binds a variable of sort $\Data$. 
\end{itemize}
We write $\CondTerm$ for the set of all closed \deACPet\ terms of sort 
$\Cond$.

Each term from $\CondTerm$ can be taken as a formula of a first-order 
language with equality of $\gD$ by taking the flexible variable 
constants as additional variables of sort $\Data$.
The flexible variable constants are implicitly taken as additional 
variables of sort $\Data$ wherever the context asks for a formula.
In this way, each term from $\CondTerm$ can be interpreted as a
formula in $\gD$.

\deACPet\ has the constants and operators of \ACPet\ and in addition the 
following operators to build terms of sort $\Proc$:
\begin{itemize}
\item
the binary \emph{guarded command} operator 
$\funct{\gc\,}{\Cond \x \Proc}{\Proc}$;
\item
for each $n \in \Nat$, for each $a \in \Act$,
the $n$-ary \emph{data parameterized action} operator
$\funct{a}
 {\underbrace{\Data \x \cdots \x \Data}_{n\; \mathrm{times}}}{\Proc}$;
\item
for each $v \in \ProgVar$, 
a unary \emph{assignment action} operator
$\funct{\assop{v}\,}{\Data}{\Proc}$;
\item
for each $\sigma \in \EvalMap$, 
a unary \emph{evaluation} operator 
$\funct{\eval{\sigma}}{\Proc}{\Proc}$.
\end{itemize}
We write $\ProcTerm$ for the set of all closed \deACPet\ terms of sort 
$\Proc$.

The same notational conventions are used as before.
Infix notation is also used for the additional binary operators.
Moreover, the notation $\ass{v}{e}$, where $v \in \ProgVar$ and $e$ is a 
\deACPet\ term of sort $\Data$, is used for the term $\assop{v}(e)$.

The notation $\phi \Liff \psi$, where $\phi$ and $\psi$ are 
\deACPet\ terms of sort $\Cond$, is used for the term
$(\phi \Limpl \psi) \Land (\psi \Limpl \phi)$.
The axioms of \deACPet\ (given below) include an equation $\phi = \psi$ 
for each two terms $\phi$ and $\psi$ from $\CondTerm$ for which the 
formula $\phi \Liff \psi$ holds in $\gD$.

Let $t$ be a term from $\ProcTerm$, $\phi$ be a term from $\CondTerm$, 
$e_1,\ldots,e_n$ and $e$ be terms from $\DataTerm$, and $a$ be a basic
action from $\Act$. 
Then the additional operators to build terms of sort $\Proc$ can be 
explained as follows:
\begin{itemize}
\item
the term $\phi \gc t$ denotes the process that behaves as the process 
denoted by $t$ if condition $\phi$ holds and as $\dead$ otherwise;
\item
the term $a(e_1,\ldots,e_n)$ denotes the process that performs the 
data parameterized action $a(e_1,\ldots,e_n)$ and after that terminates 
successfully;
\item
the term $\ass{v}{e}$ denotes the process that performs the assignment 
action $\ass{v}{e}$, whose intended effect is the assignment of the 
result of evaluating $e$ to flexible variable $v$, and after that 
terminates successfully; 
\item
the term $\eval{\sigma}(t)$ denotes the process that behaves the same as 
the process denoted by $t$ except that each subterm of $t$ that belongs 
to $\DataTerm$ is evaluated using the evaluation map $\sigma$ updated 
according to the assignment actions that have taken place at the point 
where the subterm is encountered.
\end{itemize}
Evaluation operators are a variant of state operators 
(see e.g.~\cite{BB88}).

The following closed \deACPet\ term is reminiscent of a program that 
computes the difference between two integers by subtracting the smaller 
one from the larger one ($i, j, d \in \ProgVar$):
\begin{ldispl}
\ass{d}{i} \seqc  
((d \geq j = \Btrue)  \gc \ass{d}{d - j} \altc 
 (d \geq j = \Bfalse) \gc \ass{d}{j - d})\;.%
\footnotemark
\end{ldispl}%
\footnotetext
{Here and in examples to come, the carrier of $\Data$ is assumed to be
 the set of all integers. Moreover, the usual integer constants, 
 operators  on integers, and predicates on integers are assumed
 (where operators with result sort $\Bool$ serve as predicates).}%
That is, the final value of $d$ is the absolute value of the result of
subtracting the initial value of $i$ from the initial value of $j$.
An evaluation operator can be used to show that this is the case for
given initial values of $i$ and $j$.
For example, consider the case where the initial values of $i$ and $j$
are $11$ and $3$, respectively.
Let $\sigma$ be an evaluation map such that $\sigma(i) = 11$ and
$\sigma(j) = 3$.
Then the following equation can be derived from the axioms of \deACPet\
given below:
\begin{ldispl}
\eval{\sigma}
(\ass{d}{i} \seqc  
 ((d \geq j = \Btrue)  \gc \ass{d}{d - j} \altc
  (d \geq j = \Bfalse) \gc \ass{d}{j - d})) \\
\; {} =
\ass{d}{11} \seqc \ass{d}{8}\;. 
\end{ldispl}%
This equation shows that in the case where the initial values of $i$ 
and $j$ are $11$ and $3$ the final value of $d$ is $8$ (which is the 
absolute value of the result of subtracting $11$ from $3$).

An evaluation map $\sigma$ can be extended homomorphically from flexible
variables to \deACPet\ terms of sort $\Data$ and \deACPet\ terms of sort 
$\Cond$.
These extensions are denoted by $\sigma$ as well.
Below, we write $\sigma\mapupd{e}{v}$ for the evaluation map $\sigma'$ 
defined by $\sigma'(v') = \sigma(v')$ if $v' \neq v$ and 
$\sigma'(v) = e$.

Three subsets of $\ProcTerm$ are defined:
\begin{ldispl}
\begin{aeqns}
\AProcDPA  & = & {} \Union_{n \in \Natpos}
\set{a(e_1,\dots,e_n) \where
     a \in \Act \Land e_1,\dots,e_n \in \DataTerm}\;, \\
\AProcASS & = & 
\set{\ass{v}{e} \where v \in \ProgVar \Land e \in \DataTerm}\;, \\
\AProcTerm & = & 
\set{a \where a \in \Act} \union \AProcDPA \union \AProcASS\;.
\end{aeqns}
\end{ldispl}%
In \deACPet, the elements of $\AProcTerm$ are the terms from 
$\ProcTerm$ that denote the processes that are considered to be atomic.
Henceforth, we write $\AProcTermt$ for $\AProcTerm \union \set{\tau}$ 
and $\AProcTermtd$ for $\AProcTerm \union \set{\tau,\dead}$.

The axioms of \deACPet\ are the axioms presented in 
Table~\ref{axioms-ACPet}, on the understanding that $\alpha$ now stands 
for an arbitrary term from $\AProcTermtd$,\, $H$ now stands for an 
arbitrary subset of $\AProcTerm$ or the set $\AProcTermt$, and $I$ now 
stands for an arbitrary subset of $\AProcTerm$, and in addition the 
axioms presented in Table~\ref{axioms-deACPet}.
\begin{table}[!t]
\caption{Additional axioms of \deACPet}
\label{axioms-deACPet}
\begin{eqntbl}
\begin{axcol}
e = e'         & \mif \Sat{\gD}{\fol{e = e'}}          & \axiom{IMP1} \\
\phi = \psi    & \mif \Sat{\gD}{\fol{\phi \Liff \psi}} & \axiom{IMP2} 
\eqnsep
\True \gc x = x                                      & & \axiom{GC1}  \\
\False \gc x = \dead                                 & & \axiom{GC2}  \\
\phi \gc \dead = \dead                               & & \axiom{GC3}  \\
\phi \gc (x \altc y) = \phi \gc x \altc \phi \gc y   & & \axiom{GC4}  \\
\phi \gc x \seqc y = (\phi \gc x) \seqc y            & & \axiom{GC5}  \\
\phi \gc (\psi \gc x) = (\phi \Land \psi) \gc x      & & \axiom{GC6}  \\
(\phi \Lor \psi) \gc x = \phi \gc x \altc \psi \gc x & & \axiom{GC7}  \\
(\phi \gc x) \leftm y = \phi \gc (x \leftm y)        & & \axiom{GC8}  \\
(\phi \gc x) \commm y = \phi \gc (x \commm y)        & & \axiom{GC9}  \\
x \commm (\phi \gc y) = \phi \gc (x \commm y)        & & \axiom{GC10} \\
\encap{H}(\phi \gc x) = \phi \gc \encap{H}(x)        & & \axiom{GC11} \\
\abstr{I}(\phi \gc x) = \phi \gc \abstr{I}(x)        & & \axiom{GC12} 
\eqnsep
\eval{\sigma}(\ep) = \ep                             & & \axiom{V0}   \\
\eval{\sigma}(\tau \seqc x) = \tau \seqc \eval{\sigma}(x)
                                                     & & \axiom{V1}   \\
\eval{\sigma}(a \seqc x) = a \seqc \eval{\sigma}(x)  & & \axiom{V2}   \\
\eval{\sigma}(a(e_1,\ldots,e_n) \seqc x) = 
a(\sigma(e_1),\ldots,\sigma(e_n)) \seqc \eval{\sigma}(x)
                                                     & & \axiom{V3}   \\
\eval{\sigma}(\ass{v}{e} \seqc x) = 
{\ass{v}{\sigma(e)} \seqc \eval{\sigma\mapupd{\sigma(e)}{v}}(x)} 
                                                     & & \axiom{V4}   \\
\eval{\sigma}(x \altc y) = \eval{\sigma}(x) \altc \eval{\sigma}(y)
                                                     & & \axiom{V5}   \\
\eval{\sigma}(\phi \gc y) = \sigma(\phi) \gc \eval{\sigma}(x)
                                                     & & \axiom{V6}   
\eqnsep
a(e_1,\ldots,e_n) \seqc x \commm b(e'_1,\ldots,e'_n) \seqc y = 
 {} \\ \quad
(e_1 = e'_1 \Land \ldots \Land e_n = e'_n) \gc c(e_1,\ldots,e_n) \seqc
(x \parc y)                    & \mif \commf(a,b) = c & \axiom{CM7Da} \\
a(e_1,\ldots,e_n) \seqc x \commm b(e'_1,\ldots,e'_m) \seqc y = \dead
  & \mif \commf(a,b) = \dead \;\mathrm{or}\; n \neq m & \axiom{CM7Db} \\
a(e_1,\ldots,e_n) \seqc x \commm \alpha \seqc y = \dead 
 & \mif \alpha \notin \AProcDPA & \axiom{CM7Dc} \\
\alpha \seqc x \commm a(e_1,\ldots,e_n) \seqc y = \dead 
 & \mif \alpha \notin \AProcDPA & \axiom{CM7Dd} \\
\ass{v}{e} \seqc x \commm \alpha \seqc y = \dead    & & \axiom{CM7De} \\
\alpha \seqc x \commm \ass{v}{e} \seqc y = \dead    & & \axiom{CM7Df}   
\eqnsep
\alpha \seqc (\phi \gc \tau \seqc (x \altc y) \altc \phi \gc x) = 
\alpha \seqc (\phi \gc (x \altc y))                 & & \axiom{BED}
\end{axcol}
\end{eqntbl}
\end{table}
In the latter table, 
$\phi$ and $\psi$ stand for arbitrary terms from $\CondTerm$,\,
$e$, $e_1,e_2,\ldots$, and $e'$, $e'_1,e'_2,\ldots$ stand for arbitrary 
terms from $\DataTerm$,\, 
$v$ stands for an arbitrary flexible variable from $\ProgVar$,\,
$\sigma$ stands for an arbitrary evaluation map from $\EvalMap$,\,
$a,b$, and $c$ stand for arbitrary basic actions from $\Act$, and
$\alpha$ stands for an arbitrary term from $\AProcTermtd$.

Axioms GC1--GC12 have been taken from~\cite{BB92c} (using a different 
numbering), but with the axioms with occurrences of Hoare's ternary 
counterpart of the guarded command operator (see below) replaced by 
simpler axioms.
Axioms CM7Da and CM7Db have been inspired by~\cite{BM09d}.
Axiom BED is axiom BE generalized to the current setting.
An equivalent axiomatization is obtained if axiom BED is replaced by 
the equation
$\alpha \seqc (\phi \gc \tau \seqc x) = \alpha \seqc (\phi \gc x)$.

Some earlier extensions of \ACP\ include Hoare's ternary counterpart of 
the binary guarded command operator (see e.g.~\cite{BB92c}).  
This operator can be defined by the equation
$\cond{x}{u}{y} = u \gc x \altc (\Lnot\, u) \gc y$. 
From this defining equation, it follows that  
$u \gc x = \cond{x}{u}{\dead}$.
In~\cite{GP94b}, a unary counterpart of the binary guarded command 
operator is used.
This operator can be defined by the equation $\guard{u} = u \gc \ep$.
From this defining equation, it follows that  
$u \gc x = \guard{u} \seqc x$ and also that $\guard{\True} = \ep$ and
$\guard{\False} = \dead$.

\section{\deACPet\ with Recursion}
\label{sect-deACPetr}

A closed \deACPet\ term of sort $\Proc$ denotes a process with a finite 
upper bound to the number of actions that it can perform. 
Recursion allows the description of processes without a finite upper 
bound to the number of actions that it can perform.

A \emph{recursive specification} over \deACPet\ is a set 
$\set{X_i = t_i \where i \in I}$, where $I$ is a finite set, 
each $X_i$ is a variable from $\cX$, each $t_i$ is a \deACPet\ term of 
sort $\Proc$ in which only variables from $\set{X_i \where i \in I}$ 
occur, and $X_i \neq X_j$ for all $i,j \in I$ with $i \neq j$. 
We write $\vars(E)$, where $E$ is a recursive specification over 
\deACPet, for the set of all variables that occur in $E$.
Let $E$ be a recursive specification and let $X \in \vars(E)$.
Then the unique equation $X \!= t \;\in\, E$ is called the 
\emph{recursion equation for $X$ in $E$}.

Below, recursive specifications over \deACPet\ are introduced in which 
the right-hand sides of the recursion equations are linear \deACPet\ 
terms.
The set $\LT$ of \emph{linear \deACPet\ terms} is inductively defined by 
the following rules:
\begin{itemize}
\item
$\dead \in \LT$;
\item 
if $\phi \in \CondTerm$, then $\phi \gc \ep \in \LT$;
\item
if $\phi \in \CondTerm$, $\alpha \in \AProcTermt$, and $X \in \cX$, then 
$\phi \gc \alpha \seqc X \in \LT$;
\item
if $t,t' \in \LT$, then $t \altc t' \in \LT$.
\end{itemize}
Let $t \in \LT$.
Then we refer to the subterms of $t$ that have the form $\phi \gc \ep$ 
or the form $\phi \gc \alpha \seqc X$ as the \emph{summands of} $t$.

Let $X$ be a variable from $\cX$ and
let $t$ be an \deACPet\ term in which $X$ occurs. 
Then an occurrence of $X$ in $t$ is \emph{guarded} if $t$ has a subterm 
of the form $\alpha \seqc t'$ where $\alpha \in \AProcTerm$ and $t'$ 
contains this occurrence of $X$.
An occurrence of a variable $X$ in a linear \deACPet\ term may not be 
guarded because a linear \deACPet\ term may have summands of the form 
$\phi \gc \tau \seqc X$.

A \emph{guarded linear recursive specification} over \deACPet\ is a 
recursive specification $\set{X_i = t_i \where i \in I}$ over \deACPet\ 
where each $t_i$ is a linear \deACPet\ term, and there does not exist an 
infinite sequence $i_0\;i_1\;\ldots\,$ over $I$ such that, for each 
$k \in \Nat$, there is an occurrence of $X_{i_{k+1}}$ in $t_{i_k}$ 
that is not guarded.

A \emph{linearizable recursive specification} over \deACPet\ is a 
recursive specification $\set{X_i = t_i \where i \in I}$ over \deACPet\ 
where each $t_i$ is rewritable to an \deACPet\ term $t'_i$, using the 
axioms of \deACPet\ in either direction and the equations in 
$\set{X_j = t_j \where j \in I \Land i \neq j}$ from left to right, such
that $\set{X_i = t'_i \where i \in I}$ is a guarded linear recursive 
specification over \deACPet.

A solution of a guarded linear recursive specification $E$ over 
\deACPet\ in some model of \deACPet\ is a set 
$\set{p_X \where X \in \vars(E)}$ of elements of the carrier of that 
model such that each equation in $E$ holds if, for all $X \in \vars(E)$, 
$X$ is assigned~$p_X$. 
A guarded linear recursive specification has a unique solution under the
equivalence defined in Section~\ref{sect-semantics} for \deACPet\ 
extended with guarded linear recursion.
If $\set{p_X \where X \in \vars(E)}$ is the unique solution of a guarded 
linear recursive specification $E$, then, for 
each $X \in \vars(E)$, $p_X$ is called the \emph{$X$-component} of the 
unique solution of $E$. 

\deACPet\ is extended with guarded linear recursion by adding constants 
for solutions of guarded linear recursive specifications over \deACPet\ 
and axioms concerning these additional constants.
For each guarded linear recursive specification $E$ over \deACPet\ and 
each $X \in \vars(E)$, a constant $\rec{X}{E}$ of sort $\Proc$, that 
stands for the $X$-component of the unique solution of $E$, is added to 
the constants of \deACPet.
The equation RDP (Recursive Definition Principle) and the conditional 
equation RSP (Recursive Specification Principle) given in 
Table~\ref{axioms-REC} are added to the axioms of \deACPet.
\begin{table}[!t]
\caption{Axioms for guarded linear recursion}
\label{axioms-REC}
\begin{eqntbl}
\begin{axcol}
\rec{X}{E} = \rec{t}{E} & \mif X \!= t \;\in\, E & \axiom{RDP} \\
E \Limpl X = \rec{X}{E} & \mif X \in \vars(E)    & \axiom{RSP} 
\end{axcol}
\end{eqntbl}
\end{table}
In RDP and RSP, $X$ stands for an arbitrary variable from $\cX$, 
\linebreak[2] 
$t$ stands for an arbitrary \deACPet\ term of sort $\Proc$,\, 
$E$ stands for an arbitrary guarded linear recursive specification over 
\deACPet, and 
the notation $\rec{t}{E}$ is used for $t$ with, for all 
$X \in \vars(E)$, all occurrences of $X$ in $t$ replaced by 
$\rec{X}{E}$.
Side conditions restrict what $X$, $t$ and $E$ stand for.

We write \deACPetr\ for the resulting theory.
Furthermore, we write $\ProcTermr$ for the set of all closed 
$\deACPetr$ terms of sort $\Proc$.

RDP and RSP together postulate that guarded linear recursive 
specifications over \deACPet\ have unique solutions: 
the equations $\rec{X}{E} = \rec{t}{E}$ and the conditional equations 
$E \Limpl X \!=\! \rec{X}{E}$ for a fixed $E$ express that the constants 
$\rec{X}{E}$ make up a solution of $E$ and that this solution is the 
only one, respectively.

Because conditional equational formulas must be dealt with in \deACPetr, 
it is understood that conditional equational logic is used in deriving 
equations from the axioms of \deACPetr.
A complete inference system for conditional equational logic can for
example be found in~\cite{BW90,Gog21a}.

We write $T \Ent t = t'$, where $T$ is \deACPetr\ or \deACPetrf\ 
(an extension of \deACPetr\ introduced below), to indicate that the 
equation $t = t'$ is derivable from the axioms of $T$ using a complete 
inference system for conditional equational logic.

The following closed \deACPetr\ term is reminiscent of a program that 
computes by repeated subtraction the quotient and remainder of dividing a
non-negative integer by a positive integer ($i, j, q, r \in \ProgVar$):
\begin{ldispl}
\ass{q}{0} \seqc \ass{r}{i} \seqc \rec{Q}{E}\;,  
\end{ldispl}%
where $E$ is the guarded linear recursive specification that consists of 
the following two equations ($Q,R \in \cX$):
\begin{ldispl}
Q = (r \geq j = \Btrue)  \gc \ass{q}{q + 1} \seqc R \altc 
    (r \geq j = \Bfalse) \gc \ep\;, \\
R = \True \gc \ass{r}{r - j} \seqc Q\;.
\end{ldispl}%
The final values of $q$ and $r$ are the quotient and remainder of 
dividing the initial value of $i$ by the initial value of $j$.
An evaluation operator can be used to show that this is the case for
given initial values of $i$ and $j$.
For example, consider the case where the initial values of $i$ and $j$
are $11$ and $3$, respectively.
Let $\sigma$ be an evaluation map such that $\sigma(i) = 11$ and
$\sigma(j) = 3$.
Then the following equation can be derived from the axioms of \deACPetr:
\begin{ldispl}
\eval{\sigma}
(\ass{q}{0} \seqc \ass{r}{i} \seqc \rec{Q}{E}) \\
\; {} =
\ass{q}{0} \seqc \ass{r}{11} \seqc \ass{q}{1} \seqc \ass{r}{8} \seqc 
\ass{q}{2} \seqc \ass{r}{5} \seqc \ass{q}{3} \seqc \ass{r}{2}\;. 
\end{ldispl}%
This equation shows that in the case where the initial values of $i$ 
and $j$ are $11$ and $3$ the final values of $q$ and $r$ are $3$ and 
$2$ (which are the quotient and remainder of dividing $11$ by $3$).

Below, use will be made of a reachability notion for the variables
occurring in a guarded linear recursive specification over \deACPet.

Let $E$ be a guarded linear recursive specification over \deACPet\ and  
let $X,Y \in \vars(E)$.
Then $Y$ \emph{is directly reachable from} $X$ in $E$, written 
\smash{$X \dreach{E} Y$}, if $Y$ occurs in the right-hand side of the 
recursion equation for $X$ in $E$.
We write \smash{$\reach{E}$} for the reflexive transitive closure of 
\smash{$\dreach{E}$}.

Processes with one or more cycles of $\tau$ actions are not definable
by guarded linear recursion alone, but they are definable by combining 
guarded linear recursion and abstraction.
An example is 
\begin{ldispl}
\abstr{\set{a}}
  (\rec{X}{\set{X = a \seqc Y \altc b, Y = a \seqc X \altc c}})\;.
\end{ldispl}%
The semantics of \deACPetr\ presented in Section~\ref{sect-semantics} 
identifies this with $b \altc \tau \seqc (b \altc c)$. 
However, the equation 
\begin{ldispl}
\abstr{\set{a}}
 (\rec{X}{\set{X = a \seqc Y \altc b, Y = a \seqc X \altc c}}) =
b \altc \tau \seqc (b \altc c)
\end{ldispl}%
is not derivable from the axioms of \deACPetr.
This is remedied by the addition of the equational axiom schema CFAR 
(Cluster Fair Abstraction Rule) that will be presented below.
This axiom schema makes it possible to abstract from a cycle of actions 
that are turned into the unobservable action $\tau$, by which only the 
ways out of the cycle remain.
The side condition on the equation concerned requires several notions to 
be made precise.

Let $E$ be a guarded linear recursive specification over \deACPet,
let $C \subseteq \vars(E)$, and
let $I \subseteq \AProcTerm$.
Then:
\begin{itemize}
\item 
$C$ \emph{is a cluster for $I$ in $E$} if, 
for each \deACPet\ term $\phi \gc \alpha \seqc X$ of sort $\Proc$ that 
is a summand of the right-hand side of the recursion equation for some 
$X' \in C$ in $E$, $X \in C$ only if $\phi \equiv \True$ and 
$\alpha \in I \union \set{\tau}$;%
\footnote{We write $\equiv$ for syntactic equality.}
\item
for each cluster $C$ for $I$ in $E$, the \emph{exit set of $C$ for 
$I$ in $E$}, written $\exits{I,E}(C)$, is the set of \deACPet\ terms of 
sort $\Proc$ defined by $t \in \exits{I,E}(C)$ iff $t$ is a summand of 
the right-hand side of the recursion equation for some $X' \in C$ in $E$
and one of the following holds:
\begin{itemize}
\item
$t \equiv \phi \gc \alpha \seqc Y$ for some $\phi \in \CondTerm$, 
$\alpha \in \AProcTermt$, and $Y \in \vars(E)$ such that 
$\alpha \notin I \union \set{\tau}$ or $Y \notin C$;
\item
$t \equiv \phi \gc \ep$ for some $\phi \in \CondTerm$;
\end{itemize}
\item
$C$ \emph{is a conservative cluster for $I$ in $E$} if $C$ is a cluster 
for $I$ in $E$ and, for each $X \in C$ and $Y \in \exits{I,E}(C)$, 
\smash{$X \reach{E} Y$}.
\end{itemize}

The cluster fair abstraction rule is presented in 
Table~\ref{axioms-CFAR}.
\begin{table}[!t]
\caption{Cluster fair abstraction rule}
\label{axioms-CFAR}
\begin{eqntbl}
\begin{axcol}
\tau \seqc \abstr{I}(\rec{X}{E}) = 
\tau \seqc \abstr{I}(\Altc{l=1}{n} \rec{t_l}{E}) \\ 
\mif 
\mbox{for some finite conservative cluster $C$ for $I$ in $E$,} \\ 
\phantom{\mif}
\mbox{$X \in C$ and $\exits{I,E}(C) = \set{t_1,\ldots,t_n}$}
 & & \axiom{CFAR}
\end{axcol}
\end{eqntbl}
\end{table}
In this table, 
$X$ stands for an arbitrary variable from $\cX$, 
$E$ stands for an arbitrary guarded linear recursive specification over 
\deACPet, 
$I$ stands for an arbitrary subset of $\AProcTerm$, and
$t_1,t_2,\ldots\,$ stand for arbitrary \deACPet\ terms of sort $\Proc$.
A side condition restricts what $X$, $E$, $I$, and $t_1,t_2,\ldots\,$  
stand for.

CFAR expresses that every cluster of $\tau$ actions will be exited 
sooner or later.
This is a fairness assumption made in the verification of many 
properties concerning the external behaviour of systems.

We write \deACPetrf\ for the theory \deACPetr\ extended with CFAR.

\section{Bisimulation Semantics}
\label{sect-semantics}

In this section, a structural operational semantics of \deACPetr\ is 
presented and a notion of branching bisimulation equivalence for 
\deACPetr\ based on this structural operational semantics is defined.

The structural operational semantics of \deACPetr\ consists of
\begin{itemize}
\item 
a binary \emph{conditional transition} relation \smash{$\step{\ell}$} on 
$\ProcTermr$ for each $\ell \in \EvalMap \x \AProcTermt$;
\item 
a unary \emph{successful termination} relation $\sterm{\sigma}$ on 
$\ProcTermr$ for each $\sigma \in \EvalMap$.
\end{itemize}
We write \smash{$\astep{t}{\gact{\sigma}{\alpha}}{t'}$} instead of 
\smash{$\tup{t,t'} \in {\step{\tup{\sigma,\alpha}}}$} and 
$\isterm{t}{\sigma}$ instead of $t \in {\sterm{\sigma}}$.

The relations from the structural operational semantics describe what 
the processes denoted by terms from $\ProcTermr$ are capable of doing as 
follows:
\begin{itemize}
\item
$\astep{t}{\gact{\sigma}{\alpha}}{t'}$: 
if the data values assigned to the flexible variables are as defined by 
$\sigma$, then the process denoted by $t$ has the potential to make a 
transition to the process denoted by $t'$ by performing action $\alpha$;
\item
$\isterm{t}{\sigma}$: 
if the data values assigned to the flexible variables are as defined  
by~$\sigma$, then the process denoted by $t$ has the potential to 
terminate successfully.
\end{itemize}

The relations from the structural operational semantics of \deACPetr\ 
are the smallest relations satisfying the rules given in 
Table~\ref{sos-deACPet}.%
\begin{table}[!p]
\caption{Transition rules for \deACPet}
\label{sos-deACPet}
\begin{ruletbl}
{} \\[-3ex]
\Rule
{}
{\astep{\alpha}{\gact{\sigma}{\alpha}}{\ep}}
\\[-2ex]
\Rule
{\phantom{\isterm{\ep}{\sigma}}}
{\isterm{\ep}{\sigma}}
\\
\Rule
{\isterm{x}{\sigma}}
{\isterm{x \altc y}{\sigma}}
\quad\;
\Rule
{\isterm{y}{\sigma}}
{\isterm{x \altc y}{\sigma}}
\quad\;
\Rule
{\astep{x}{\gact{\sigma}{\alpha}}{x'}}
{\astep{x \altc y}{\gact{\sigma}{\alpha}}{x'}}
\quad\;
\Rule
{\astep{y}{\gact{\sigma}{\alpha}}{y'}}
{\astep{x \altc y}{\gact{\sigma}{\alpha}}{y'}}
\\
\Rule
{\isterm{x}{\sigma},\; \isterm{y}{\sigma}}
{\isterm{x \seqc y}{\sigma}}
\quad\;
\Rule
{\isterm{x}{\sigma},\; \astep{y}{\gact{\sigma}{\alpha}}{y'}}
{\astep{x \seqc y}{\gact{\sigma}{\alpha}}{y'}}
\quad\;
\Rule
{\astep{x}{\gact{\sigma}{\alpha}}{x'}}
{\astep{x \seqc y}{\gact{\sigma}{\alpha}}{x' \seqc y}}
\\
\Rule
{\isterm{x}{\sigma},\; \isterm{y}{\sigma}}
{\isterm{x \parc y}{\sigma}}
\quad\;
\Rule
{\astep{x}{\gact{\sigma}{\alpha}}{x'}}
{\astep{x \parc y}{\gact{\sigma}{\alpha}}{x' \parc y}}
\quad\;
\Rule
{\astep{y}{\gact{\sigma}{\alpha}}{y'}}
{\astep{x \parc y}{\gact{\sigma}{\alpha}}{x \parc y'}}
\\
\RuleC
{\astep{x}{\gact{\sigma}{a}}{x'},\; \astep{y}{\gact{\sigma}{b}}{y'}}
{\astep{x \parc y}{\gact{\sigma}{c}}{x' \parc y'}}
{\commf(a,b) = c}
\\
\RuleC
{\astep{x}{\gact{\sigma}{a(e_1,\ldots,e_n)}}{x'},\; 
 \astep{y}{\gact{\sigma}{b(e'_1,\ldots,e'_n)}}{y'}}
{\astep{x \parc y}{\gact{\sigma}{c(e_1,\ldots,e_n)}}{x' \parc y'}}
{\commf(a,b) = c,\;
 \Sat{\gD}{\sigma(\fol{e_1 = e'_1 \Land \ldots \Land  e_n = e'_n})}}
\\
\Rule
{\astep{x}{\gact{\sigma}{\alpha}}{x'}}
{\astep{x \leftm y}{\gact{\sigma}{\alpha}}{x' \parc y}}
\\
\RuleC
{\astep{x}{\gact{\sigma}{a}}{x'},\; \astep{y}{\gact{\sigma}{b}}{y'}}
{\astep{x \commm y}{\gact{\sigma}{c}}{x' \parc y'}}
{\commf(a,b) = c}
\\
\RuleC
{\astep{x}{\gact{\sigma}{a(e_1,\ldots,e_n)}}{x'},\; 
 \astep{y}{\gact{\sigma}{b(e'_1,\ldots,e'_n)}}{y'}}
{\astep{x \commm y}
  {\gact{\sigma}{c(e_1,\ldots,e_n)}}{x' \parc y'}}
{\commf(a,b) = c,\;
 \Sat{\gD}{\sigma(\fol{e_1 = e'_1 \Land \ldots \Land  e_n = e'_n})}}
\\
\Rule
{\isterm{x}{\sigma}}
{\isterm{\encap{H}(x)}{\sigma}}
\quad\;
\RuleC
{\astep{x}{\gact{\sigma}{\alpha}}{x'}}
{\astep{\encap{H}(x)}{\gact{\sigma}{\alpha}}{\encap{H}(x')}}
{\alpha \notin H}
\\
\Rule
{\isterm{x}{\sigma}}
{\isterm{\abstr{I}(x)}{\sigma}}
\quad\;
\RuleC
{\astep{x}{\gact{\sigma}{\alpha}}{x'}}
{\astep{\abstr{I}(x)}{\gact{\sigma}{\alpha}}{\abstr{I}(x')}}
{\alpha \notin I}
\quad\;
\RuleC
{\astep{x}{\gact{\sigma}{\alpha}}{x'}}
{\astep{\abstr{I}(x)}{\gact{\sigma}{\tau}}{\abstr{I}(x')}}
{\alpha \in I}
\\
\RuleC
{\isterm{x}{\sigma}}
{\isterm{\phi \gc x}{\sigma}}
{\Sat{\gD}{\sigma(\fol{\phi})}}
\quad\;
\RuleC
{\astep{x}{\gact{\sigma}{\alpha}}{x'}}
{\astep{\phi \gc x}{\gact{\sigma}{\alpha}}{x'}}
{\Sat{\gD}{\sigma(\fol{\phi})}}
\\
\Rule
{\isterm{x}{\sigma}}
{\isterm{\eval{\sigma}(x)}{\sigma'}}
\quad\;
\Rule
{\astep{x}{\gact{\sigma}{\tau}}{x'}}
{\astep{\eval{\sigma}(x)}{\gact{\sigma'}{\tau}}{\eval{\sigma}(x')}}
\quad\;
\Rule
{\astep{x}{\gact{\sigma}{a}}{x'}}
{\astep{\eval{\sigma}(x)}{\gact{\sigma'}{a}}{\eval{\sigma}(x')}}
\\
\Rule
{\astep{x}{\gact{\sigma}{a(e_1,\ldots,e_n)}}{x'}}
{\astep{\eval{\sigma}(x)}
       {\gact{\sigma'}{a(\sigma(e_1),\ldots,\sigma(e_n))}}
       {\eval{\sigma}(x')}}
\quad\;
\Rule
{\astep{x}{\gact{\sigma}{\ass{v}{e}}}{x'}}
{\astep{\eval{\sigma}(x)}{\gact{\sigma'}{\ass{v}{\sigma(e)}}}
       {\eval{\sigma\mapupd{\sigma(e)}{v}}(x')}}
\\
\RuleC
{\isterm{\rec{t}{E}}{\sigma}}
{\isterm{\rec{X}{E}}{\sigma}}
{X \!\!=\! t \,\in\, E}
\qquad
\RuleC
{\astep{\rec{t}{E}}{\gact{\sigma}{\alpha}}{x'}}
{\astep{\rec{X}{E}}{\gact{\sigma}{\alpha}}{x'}}
{X \!\!=\! t \,\in\, E}
\vspace*{1ex}
\end{ruletbl}
\end{table}
In this table, 
$\sigma$ and $\sigma'$ stand for arbitrary evaluation maps from 
$\EvalMap$,\,
$\alpha$ stands for an arbitrary action from~$\AProcTermt$,\,
$a,b$, and $c$ stand for arbitrary actions from $\Act$,\,
$e,e_1,e_2,\ldots$ and $e'_1,e'_2,\ldots$ stand for arbitrary terms 
from~$\DataTerm$,\,
$H$ stands for an arbitrary subset of $\AProcTerm$ or the set 
$\AProcTermt$,\, 
$I$ stands for an arbitrary subset of~$\AProcTerm$,\,
$\phi$ stands for an arbitrary term from $\CondTerm$,\,
$v$ stands for an arbitrary flexible variable from $\ProgVar$,\,
$X$ stands for an arbitrary variable from $\cX$,\, 
$t$ stands for an arbitrary \deACPet\ term of sort $\Proc$, and
$E$ stands for an arbitrary guarded linear recursive specification over 
\deACPet.

The rules in Table~\ref{sos-deACPet} have the form 
\smash{\small$\SRuleC{p_1,\ldots,p_n}{\raisebox{.6ex}{$c$}}{s}$}, where 
$s$ is optional.
They are to be read as ``if $p_1$ and \ldots and $p_n$ then $c$, 
provided $s$''.
As usual, $p_1,\ldots,p_n$ are called the premises and $c$ is called the 
conclusion.
A side condition $s$, if present, serves to restrict the applicability 
of a rule.
If a rule has no premises, then nothing is displayed above the 
horizontal bar.

\sloppy
Because the rules in Table~\ref{sos-deACPet} constitute an inductive 
definition, \mbox{$\astep{t}{\gact{\sigma}{\alpha}}{t'}$} or 
$\isterm{t}{\sigma}$ holds iff it can be inferred from these rules.
For instance, for \mbox{$a,b,c \in \Act$}, $v,v' \in \ProgVar$, and 
$\sigma \in \EvalMap$ such that $\sigma(v) = \sigma(v')$, we have that
\smash{$\astep{(v = v') \gc (a \altc b) \seqc c}{\gact{\sigma}{a}}
              {\ep \seqc c}$}
can be inferred by applying the first rule, the third rule for $\altc$,
the second rule for $\gc$, and the third rule for $\seqc$ in that order.

Two processes are considered equal if they can simulate each other 
insofar as their observable potentials to make transitions and to 
terminate successfully are concerned, taking into account the assigments 
of data values to flexible variables under which the potentials are 
available.
This can be dealt with by means of the notion of branching bisimulation 
equivalence introduced in~\cite{GW96a} adapted to the conditionality of 
transitions in which the unobservable action $\tau$ is performed.

An equivalence relation on the set $\AProcTermt$ is needed.
Two actions $\alpha,\alpha' \in \AProcTermt$ are \emph{data equivalent}, 
written $\alpha \simeq \alpha'$, iff one of the following holds:
\begin{itemize}
\item
there exists an $a \in \Actt$ such that $\alpha = a$ 
and $\alpha' = a$;
\item
for some $n \in \Natpos$,
there exist an $a \in \Act$ and 
$e_1,\dots,e_n,e'_1,\dots,e'_n \in \DataTerm$
such that 
$\Sat{\gD}{\fol{e_1 = e'_1}}$, \ldots, $\Sat{\gD}{\fol{e_n = e'_n}}$,
$\alpha = a(e_1,\dots,e_n)$, and $\alpha' = a(e'_1,\dots,e'_n)$;
\item
there exist a $v \in \ProgVar$ and $e,e' \in \DataTerm$ such that 
$\Sat{\gD}{\fol{e = e'}}$, $\alpha = \ass{v}{e}$, and 
$\alpha' = \ass{v}{e'}$.
\end{itemize}
We write $[\alpha]$, where $\alpha \in \AProcTermt$, for the equivalence 
class of $\alpha$ with respect to $\simeq$.

For each $\sigma \in \EvalMap$, the binary relation $\silent{\sigma}$ on 
$\ProcTermr$ defined as the reflexive transitive closure of 
\smash{$\step{\gact{\sigma}{\tau}}$} is also needed.

Moreover, we write $\astep{t}{(\gact{\sigma}{\alpha})}{t'}$, 
where $\sigma \in \EvalMap$ and $\alpha \in \AProcTermt$, for 
\smash{$\astep{t}{\gact{\sigma}{\alpha}}{t'}$} or both $\alpha = \tau$ 
and $t = t'$.

A \emph{branching bisimulation} is a binary relation $R$ on 
$\ProcTermr$ such that, for all terms $t_1,t_2 \in \ProcTermr$ with 
$(t_1,t_2) \in R$, the following \emph{transfer conditions} hold:
\begin{itemize}
\item
if $\astep{t_1}{\gact{\sigma}{\alpha}}{t_1'}$, then there exist an 
$\alpha' \in [\alpha]$ and $t_2',t_2^* \in \ProcTermr$ such that 
$t_2 \silent{\sigma} t_2^*$, 
$\astep{t_2^*}{(\gact{\sigma}{\alpha'})}{t_2'}$, 
$(t_1,t_2^*) \in R$, and $(t_1',t_2') \in R$;
\item
if $\astep{t_2}{\gact{\sigma}{\alpha}}{t_2'}$, then there exist an 
$\alpha' \in [\alpha]$ and $t_1',t_1^* \in \ProcTermr$ such that 
$t_1 \silent{\sigma} t_1^*$, 
$\astep{t_1^*}{(\gact{\sigma}{\alpha'})}{t_1'}$, 
$(t_1^*,t_2) \in R$, and $(t_1',t_2') \in R$;
\item
if $\isterm{t_1}{\sigma}$, then there exists a $t_2^* \in \ProcTermr$ 
such that $t_2 \silent{\sigma} t_2^*$, $\isterm{t_2^*}{\sigma}$, and 
\mbox{$(t_1,t_2^*) \in R$};
\item
if $\isterm{t_2}{\sigma}$, then there exists a $t_1^* \in \ProcTermr$ 
such that $t_1 \silent{\sigma} t_1^*$, $\isterm{t_1^*}{\sigma}$, and 
\mbox{$(t_1^*,t_2) \in R$}.
\end{itemize}
If $R$ is a branching bisimulation, then a pair $(t_1,t_2)$ is said to
satisfy the \emph{root condition} in $R$ if the following conditions 
hold:
\begin{itemize}
\item
if $\astep{t_1}{\gact{\sigma}{\alpha}}{t_1'}$, then there exist an 
$\alpha' \in [\alpha]$ and a $t_2' \in \ProcTermr$ such that 
$\astep{t_2}{\gact{\sigma}{\alpha'}}{t_2'}$ and $(t_1',t_2') \in R$;
\item
if $\astep{t_2}{\gact{\sigma}{\alpha}}{t_2'}$, then there exist an 
$\alpha' \in [\alpha]$ and a $t_1' \in \ProcTermr$ such that 
$\astep{t_1}{\gact{\sigma}{\alpha'}}{t_1'}$ and $(t_1',t_2') \in R$;
\item
$\isterm{t_1}{\sigma}$ iff $\isterm{t_2}{\sigma}$.
\end{itemize}

Two terms $t_1,t_2 \in \ProcTermr$ are 
\emph{rooted branching bisimulation equivalent}, 
written $t_1 \rbbisim t_2$, if there exists a branching bisimulation $R$ 
such that $(t_1,t_2) \in R$ and $(t_1,t_2)$ satisfies the root condition 
in $R$.

In Section~\ref{sect-sound-compl}, it is proved that $\rbbisim$ is a 
congruence with respect to the operators of \deACPetr\ of which the 
result sort and at least one argument sort is $\Proc$.
Without the root condition, $\rbbisim$ would not be a congruence with 
respect to the operator $\altc$.
For example, it would be the case that $\tau \seqc a \rbbisim a$ and not 
$\tau \seqc a \altc b \rbbisim a \altc b$.

Let $R$ be a branching bisimulation such that $(t_1,t_2) \in R$ and the 
pair $(t_1,t_2)$ satisfies the root condition in $R$.
Then we say that $R$ is a branching bisimulation \emph{witnessing}
$t_1 \rbbisim t_2$.

\section{Interlude}
\label{sect-interlude}

In the preceding sections, two relatively uncommon choices have been 
made:
\begin{itemize}
\item
the choice to include the rather unusual evaluation operators, i.e.\ 
$\eval{\sigma}$ for each $\sigma \in \EvalMap$, in the operators of 
\deACPetr;
\item
the choice for a structural operational semantics of \deACPetr\ with a 
transition relation \smash{$\step{\ell}$} on $\ProcTermr$ for each 
$\ell \in \EvalMap \x \AProcTermt$, while a transition relation 
\smash{$\step{\ell}$} on $\ProcTermr \x \EvalMap$ for each 
$\ell \in \AProcTermt$ is arguably more common.
\end{itemize}
In this short section, the reasons for these choices are clarified.

The issues which influenced the above-mentioned choices most are:
\begin{itemize}
\item
the need for a variant of rooted branching bisimulation equivalence
that is a congruence with respect to all operators of \deACPetr;
\item
the need for a coarser equivalence in cases where parallel composition, 
left merge, and communication merge are not involved.
\end{itemize}

With the chosen kind of transition relations, the first need can be 
fulfilled with a simple and natural generalization of rooted branching 
bisimulation equivalence as originally introduced in~\cite{GW96a}, but 
with the more common kind of transition relations a less obvious variant 
of rooted branching bisimulation equivalence, a `stateless' variant in 
the terminology of~\cite{MRG05a}, has to be devised.
As a consequence, with the chosen kind of transition relations, 
generalizations of existing proof techniques and proof ideas could be 
used in establishing the soundness and semi-completeness results 
presented in Section~\ref{sect-sound-compl}, whereas this would not be 
the case with the more common kind of transition relations.

The variant of rooted branching bisimulation equivalence referred to at 
the beginning of the previous paragraph is the equivalence $\rbbisim$ 
introduced at the end of Section~\ref{sect-semantics}. 
In order to be a congruence with respect to parallel com\-position, 
left merge, and communication merge, $\rbbisim$ identifies two terms 
from $\ProcTermr$ if the processes denoted by them can simulate each 
other even in the case where the data values assigned to flexible 
variables may change after each transition --- through assignment actions 
performed by parallel processes.
The second need mentioned above is the need for an equivalence that does
not takes such changes into account in cases where parallel composition, 
left merge, and communication merge are not involved.
Two terms $t,t' \in \ProcTermr$ are equivalent according to this 
coarser equivalence iff $\eval{\sigma}(t) \rbbisim \eval{\sigma}(t')$ 
for all $\sigma \in \EvalMap$.
This means that the coarser equivalence is covered in the semantics of 
\deACPetr\ by the choice to include the evaluation operators in the 
operators of \deACPetr.

We have that, 
for all $t,t' \in \ProcTermr$ and $\sigma \in \EvalMap$, 
$\deACPetrf \Ent \eval{\sigma}(t) = \eval{\sigma}(t')$ iff 
$\eval{\sigma}(t) \rbbisim \eval{\sigma}(t')$ 
(see Corollary~\ref{corollary-completeness-ACPet} below).
So the axioms of \deACPetrf, which constitute an equational 
axiomatization of $\rbbisim$, are also adequate for equational 
verification of the coarser equivalence.

In~\cite{GP94b}, an extension of \ACP\ with the empty process constant, 
the unary counterpart of the binary guarded command operator, and 
actions to change a data-state is presented. 
Evaluation maps can be taken as special cases of data-states.
For similar reasons as in the case of \deACPetr, there is a need for
two equivalences.
This is not dealt with by the inclusion of evaluation operators.
Instead, in equational reasoning, certain axioms may only be applied to
terms in which the parallel composition, left merge, and communication 
merge operators do not occur.

In an appendix, a structural operational semantics of \deACPetr\ is 
presented which is reminiscent of a symbolic operational semantics in 
the sense of~\cite{HL95a}.
It is a structural operational semantics with a transition relation 
\smash{$\step{\ell}$} on $\ProcTermr$ for each 
$\ell \in \sCondTerm \x \AProcTermt$, where $\sCondTerm$ is the set of 
all terms $\phi \in \CondTerm$ for which 
$\nSat{\gD}{\phi \Liff \False}$.
In my opinion, this structural operational semantics is intuitively more 
appealing than the one presented in Section~\ref{sect-semantics}, but 
the definition of the variant of rooted branching bisimulation 
equivalence based on it is quite unintelligible.

\section{Soundness and Completeness}
\label{sect-sound-compl}

In this section, soundness and (semi-)completeness results with respect 
\sloppy to branching bisimulation equivalence for the axioms of 
\deACPetrf\ are presented.

Firstly, rooted branching bisimulation equivalence is an equivalence 
relation indeed.%
\begin{proposition}[Equivalence]
\label{proposition-equiv-deACPet}
The relation $\rbbisim$ is an equivalence relation. 
\end{proposition}
\begin{proof}
It must be shown that $\rbbisim$ is reflexive, symmetric, and 
transitive.

Let $t \in \ProcTermr$.
Then the identity relation $I$ on $\ProcTermr$ is a branching 
bisim\-ulation such that $(t,t) \in I$ and $(t,t)$ satisfies the root 
condition in $I$.
Hence, $t \rbbisim t$, which proves that $\rbbisim$ is reflexive.

Let $t_1, t_2 \in \ProcTermr$ be such that $t_1 \rbbisim t_2$, and
let $R$ be a branching bisimulation such that $(t_1,t_2) \in R$ and 
$(t_1,t_2)$ satisfies the root condition in $R$.
Then $R^{-1}$ is a branching bisimulation such that 
$(t_2,t_1) \in R^{-1}$ and $(t_2,t_1)$ satisfies the root condition in 
$R^{-1}$.
Hence, $t_2 \rbbisim t_1$, which proves that $\rbbisim$ is symmetric.

Let $t_1, t_2, t_3 \in \ProcTermr$ be such that $t_1 \rbbisim t_2$ and
$t_2 \rbbisim t_3$, 
let $R$ be a branching bisimulation such that $(t_1,t_2) \in R$ and 
$(t_1,t_2)$ satisfies the root condition in $R$, and 
let $S$ be a branching bisimulation such that $(t_2,t_3) \in S$ and 
$(t_2,t_3)$ satisfies the root condition in $S$.
Then $R \circ S$ is a branching bisimulation such that 
$(t_1,t_3) \in R \circ S$ and $(t_1,t_3)$ satisfies the root condition 
in $R \circ S$.%
\footnote
{We write $R \circ S$ for the composition of $R$ with $S$.}
That $R \circ S$ is a branching bisimulation is proved in the same way 
as Proposition~7 in~\cite{Bas96a}.
Hence, $t_1 \rbbisim t_3$, which proves that $\rbbisim$ is transitive.
\qed
\end{proof}

Moreover, rooted branching bisimulation equivalence is a congruence with 
respect to the operators of \deACPetr\ of which the result sort and at 
least one argument sort is $\Proc$.
\begin{proposition}[Congruence]
\label{proposition-congr-deACPet}
For all terms $t_1,t_1',t_2,t_2' \in \ProcTermr$ and all terms 
$\phi \in \CondTerm$, 
$t_1 \rbbisim t_2$ and $t_1' \rbbisim t_2'$ only if 
$t_1 \altc t_1' \rbbisim t_2 \altc t_2'$, 
$t_1 \seqc t_1' \rbbisim t_2 \seqc t_2'$, 
$t_1 \parc t_1' \rbbisim t_2 \parc t_2'$, 
$t_1 \leftm t_1' \rbbisim t_2 \leftm t_2'$,
$t_1 \commm t_1' \rbbisim t_2 \commm t_2'$,
$\encap{H}(t_1) \rbbisim \encap{H}(t_2)$, 
$\abstr{I}(t_1) \rbbisim \abstr{I}(t_2)$,
$\phi \gc t_1 \rbbisim \phi \gc t_2$, and
$\eval{\sigma}(t_1) \rbbisim \eval{\sigma}(t_2)$.
\end{proposition}
\begin{proof}
A detailed proof would contain an adapted copy of at least ten pages 
from~\cite{Fok00a}.
Therefore, only an outline of the proof is given here.
In order to fully understand the outline, the above-mentioned paper must 
be consulted.   

In~\cite{Fok00a}, an SOS rule format is presented which guarantees that 
the `standard' version of branching bisimulation equivalence is a 
congruence.
The format concerned is called the RBB safe format.
Below, this format is adapted in order to deal with a set of transition 
labels that contains a special element $\gact{\sigma}{\tau}$ for each 
$\sigma \in \EvalMap$ instead of a single special element $\tau$ and 
with a slightly different version of branching bisimulation equivalence.
A definition of a patience rule is needed that differs from the one 
given in~\cite{Fok00a}: a \emph{patience rule} for the $i$th argument of 
an $n$-ary operator $f$ is a path rule of the form
\begin{ldispl}
\SRule
{\astep{x_i}{\gact{\sigma}{\tau}}{y}}
{\astep{f(x_1,\ldots,x_{i-1},x_i,x_{i+1},\ldots,x_n)}
       {\gact{\sigma}{\tau}}
       {f(x_1,\ldots,x_{i-1},y,x_{i+1},\ldots,x_n)}}\;,
\end{ldispl}%
where $\sigma \in \EvalMap$.
The RBB safe format is adapted by making the following changes to the 
definition of the RBB safe format as given in~\cite{Fok00a}: 
\begin{itemize} 
\item
in the two syntactic restrictions of the RBB safe format that concern 
wild arguments, the phrase ``a patience rule'' is changed to 
``a patience rule for each $\sigma \in \EvalMap$'';
\item
in the second syntactic restrictions of the RBB safe format that concern 
wild arguments, the phrase ``the relation $\step{\tau}$'' is changed to 
``the relation \smash{$\step{\gact{\sigma}{\tau}}$} for some 
$\sigma \in \EvalMap$''.
\end{itemize} 
It is straightforward to check that the proof of Theorem~3.4 
from~\cite{Fok00a} goes through for the adapted RBB safe format and the 
version of branching bisimulation equivalence considered in this paper.
This means that the proposition holds if the rules in 
Table~\ref{sos-deACPet} are in the adapted RBB safe format with respect 
to some tame/wild labeling of arguments of operators.
It is easy to verify that this is the case with the following tame/wild 
labeling:
both arguments of $\altc$ are tame, 
the first argument of $\seqc$ is wild and 
the second argument of $\seqc$ is tame,
both arguments of $\parc$ are wild,
both arguments of $\leftm$ and $\commm$ are tame, 
the argument of $\encap{H}$ and $\abstr{I}$ is wild,
the second argument of $\gc$ is tame, and
the argument of $\eval{\sigma}$ is wild.
\qed
\end{proof}

\noindent
The tame/wild labeling given at the end of the proof of 
Proposition~\ref{proposition-congr-deACPet} is provided so that the 
reader who consults~\cite{Fok00a} can easily check that the rules in 
Table~\ref{sos-deACPet} are in the adapted RBB safe format.

Below, the soundness of the axiom system of \deACPetrf\ with respect to 
${\rbbisim}$ for equations between terms from $\ProcTermr$ will 
be established.

The following terminology will be used in the soundness proof: 
(a)~an equation $\eqn$ of \deACPetr\ terms of sort $\Proc$ is said to be 
\emph{valid with respect to} ${\rbbisim}$ if, for each closed 
substitution instance $t = t'$ of $\eqn$, $t \rbbisim t'$ and
(b)~a conditional equation $\ceqn$ of \deACPetr\ terms of sort $\Proc$ 
is said to be \emph{valid with respect to} ${\rbbisim}$ if, for each 
closed substitution instance 
$\set{t_i = t'_i \where i \in I} \Limpl t = t'$ 
of $\ceqn$, $t \rbbisim t'$ if $t_i \rbbisim t'_i$ for each $i \in I$.

\begin{theorem}[Soundness]
\label{theorem-soundness-ACPet}
For all terms $t,t' \in \ProcTermr$, $t = t'$ is derivable from the 
axioms of \deACPetrf\ only if $t \rbbisim t'$.
\end{theorem}
\begin{proof}
Because ${\rbbisim}$ is a congruence with respect to all operators from 
the signature of \deACPetrf, only the validity of each axiom of 
\deACPetrf\ has to be proved.

Below, we write $\csi(\eqn)$, where $\eqn$ is an equation of \deACPetr\ 
terms of sort $\Proc$, for the set of all closed substitution instances 
of $\eqn$.
Moreover, we write $R_\id$ for the identity relation on $\ProcTermr$.

For each axiom $\ax$ of \deACPetrf, a rooted branching bisimulation 
$R_\ax$ witnessing the validity of $\ax$ can be constructed as follows:
\begin{itemize}
\item
if $\ax$ is one of the axioms A7, CM2E, CM5E, CM6E, GC2 or an instance 
of one of the axiom schemas D0, D2, T0, GC3, V0, CM7Db--CM7Df:
\begin{ldispl}
R_\ax = \set{\tup{t,t'} \where t = t' \in \csi(\ax)}\;;
\end{ldispl}%
\item
if $\ax$ is one of the axioms A1--A6, A8, A9, CM4, CM8--CM9, GC1 or an 
instance of one of the axiom schemas CM3, CM7, D1, D3, D4, T1--T4,  
GC4--GC12, V1--V6, CM7Da, RDP:
\begin{ldispl}
R_\ax = \set{\tup{t,t'} \where t = t' \in \csi(\ax)} \union R_\id\;;
\end{ldispl}%
\item
if $\ax$ is CM1E:
\begin{ldispl}
R_\ax = 
\set{\tup{t,t'} \where t = t' \in \csi(\ax)}
\\ \phantom{R_\ax = {}}\,
 {} \union
\set{\tup{t,t'} \where t = t' \in \csi(x \parc y = y \parc x)} \union 
R_\id\;;
\end{ldispl}%
\item
if $\ax$ is an instance of BE:
\begin{ldispl}
R_\ax = 
\set{\tup{t,t'} \where t = t' \in \csi(\ax)} 
\\ \phantom{R_\ax = {}}\,
 {} \union
 \set{\tup{t,t'} \where 
      t = t' \in \csi(\tau \seqc (x \altc y) \altc x = x \altc y)}
\union R_\id\;;
\end{ldispl}%
\item
if $\ax$ is an instance of BED: similar;
\item
if $\ax$ is an instance 
$\tau \seqc \abstr{I}(\rec{X}{E}) = 
 \tau \seqc \abstr{I}(\Altc{l=1}{n} \rec{t_l}{E})$
of CFAR:
\begin{ldispl}
R_\ax = 
\set{\tup{\tau \seqc \abstr{I}(\rec{X}{E}), 
          \tau \seqc \abstr{I}(\Altc{l=1}{n} \rec{t_l}{E})}} 
\\ \phantom{R_\ax = {}}\, 
 {} \union
\set{\tup{\abstr{I}(\rec{X'}{E}), 
          \abstr{I}(\Altc{l=1}{n} \rec{t_l}{E})} \where X' \in C}
 \union R_\id\;,
\end{ldispl}%
where $C$ is the finite conservative cluster for $I$ in $E$ such that
$X \in C$ and $\exits{I,E}(C) = \set{t_1,\ldots,t_n}$; 
\item
if $\ax$ is an instance 
$\set{X_i = t_i \where i \in I} \Limpl 
 X_j = \rec{X_j}{\set{X_i = t_i \where i \in I}}$ 
($j \in I$) of RSP:
\begin{ldispl}
R_\ax = 
\set{\tup{\theta(X_j),\rec{X_j}{\set{X_i = t_i \where i \in I}}} \where 
\\ \phantom{R_\ax = {}\{}\, 
     j \in I \Land \theta \in \Theta \Land
     \LAND_{i \in I} \theta(X_i) \rbbisim \theta(t_i)}
 \union R_\id\;,
\end{ldispl}%
where 
$\Theta$ is the set of all functions from $\cX$ to $\ProcTermr$ and 
$\theta(t)$, where $\theta \in \Theta$ and $t \in \ProcTermr$, stands
for $t$ with, for all $X \in \cX$, all occurrences of $X$ replaced by 
$\theta(X)$.
\end{itemize}
For each equational axiom $\ax$ of \deACPetrf, it is straightforward to 
check that the constructed relation $R_\ax$ is a branching bisimulation 
witnessing, for each closed substitution instance $t = t'$ of $\ax$, 
$t \rbbisim t'$.
For each conditional equational axiom $\ax$ of \deACPetrf, i.e.\ for 
each instance of RSP, it is straightforward to check that the 
constructed relation $R_\ax$ is a branching bisimulation witnessing, for 
each closed substitution instance 
$\set{t_i = t'_i \where i \in I} \Limpl t = t'$ 
of $\ax$, $t \rbbisim t'$ if $t_i \rbbisim t'_i$ for each $i \in I$.
\qed
\end{proof}

The axioms of \deACPetrf\ are incomplete with respect to $\rbbisim$ 
for equations between terms from $\ProcTermr$ and there is no 
straightforward way to rectify this.
Below two semi-completeness results are presented.
The next two lemmas are used in the proofs of those results.

A term $t \in \ProcTermr$ is called \emph{abstraction-free} if no 
abstraction operator occurs in~$t$.
A term $t \in \ProcTermr$ is called \emph{bool-conditional} if,
for each  $\phi \in \CondTerm$ that occurs in~$t$, 
$\Sat{\gD}{\fol{\phi \Liff \True}}$ or 
$\Sat{\gD}{\fol{\phi \Liff \False}}$.

\begin{lemma}
\label{lemma-glr-abstr-free}
For all abstraction-free $t \in \ProcTermr$, there exists a guarded
lin\-ear recursive specification $E$ and $X \in \vars(E)$ such that
$\deACPetr \Ent \linebreak[2] t = \rec{X}{E}$.
\end{lemma}
\begin{proof}
This is easily proved by structural induction on $t$.
The proof involves constructions of guarded linear recursive 
specifications from guarded linear recursive specifications for the 
operators of \deACPet\ other than the abstraction operators.
For the greater part, the constructions are reminiscent of operations 
on process graphs defined in Sections~2.7 and~4.5.5 from~\cite{BW90}.
\qed
\end{proof}

\begin{lemma}
\label{lemma-glr-bool-cond}
\sloppy
For all bool-conditional $t \in \ProcTermr$, there exists a guarded
linear recursive specification $E$ and $X \in \vars(E)$ such that
$\deACPetrf \Ent \mbox{$t = \rec{X}{E}$}$.
\end{lemma}
\begin{proof}
This is also proved by structural induction on $t$.
The cases other than the case where $t$ is of the form $\abstr{I}(t')$
are as in the proof of Lemma~\ref{lemma-glr-abstr-free}.
The case where $t$ is of the form $\abstr{I}(t')$ is the difficult one.
It is proved in the same way as it is done for \ACPtrf\ in the proof of 
Theorem~5.6.2 from~\cite{Fok00}.
\qed
\end{proof}

\noindent
The difficult case of the proof of Lemma~\ref{lemma-glr-bool-cond} is 
the only case in which an application of CFAR is involved.

The following two theorems are the semi-completeness results referred to 
above.
\begin{theorem}[Semi-completeness I]
\label{theorem-completeness-ACPet-1}
\sloppy
For all abstraction-free $t,t' \in \ProcTermr$, 
$\deACPetr \Ent \mbox{$t = t'$}$ if $t \rbbisim t'$.
\end{theorem}
\begin{proof}
Because of Lemma~\ref{lemma-glr-abstr-free}, 
Theorem~\ref{theorem-soundness-ACPet}, and
Proposition~\ref{proposition-equiv-deACPet}, it suffices to prove that, 
for all guarded linear recursive specifications $E$ and $E'$ with 
\mbox{$X \in \vars(E)$} and $X' \in \vars(E')$, 
$\deACPetr \Ent \rec{X}{E} = \rec{X'}{E'}$ if 
$\rec{X}{E} \rbbisim \rec{X'}{E'}$. 
This is proved in the same way as it is done for \ACPtr\ in the proof 
of Theorem~5.3.2 from~\cite{Fok00}.
\qed
\end{proof}

\begin{theorem}[Semi-completeness II]
\label{theorem-completeness-ACPet-2}
For all bool-conditional $t,t' \in \ProcTermr$, 
$\deACPetrf \Ent \mbox{$t = t'$}$ if $t \rbbisim t'$.
\end{theorem}
\begin{proof}
Because of Lemma~\ref{lemma-glr-bool-cond}, 
Theorem~\ref{theorem-soundness-ACPet}, and
Proposition~\ref{proposition-equiv-deACPet}, it suffices to prove that, 
for all guarded linear recursive specifications $E$ and $E'$ with 
$X \in \vars(E)$ and $X' \in \vars(E')$, 
$\deACPetrf \Ent \rec{X}{E} = \rec{X'}{E'}$ if 
$\rec{X}{E} \rbbisim \nolinebreak \rec{X'}{E'}$. 
This is proved in the same way as it is done for \ACPtr\ in the proof 
of Theorem~5.3.2 from~\cite{Fok00}.
\qed
\end{proof}

\noindent
It is due to sufficiently similar shapes of linear \deACPet\ terms and 
linear \ACPt\ terms that parts of the proof of 
Theorems~\ref{theorem-completeness-ACPet-1} 
and~\ref{theorem-completeness-ACPet-2} go in the same way as parts of 
proofs from~\cite{Fok00}.
It needs mentioning here that, the body of the proof of Theorem~5.3.2 
from~\cite{Fok00} is restricted to constants $\rec{X}{E}$ where $E$ does 
not contain equations $Y = \tau \altc \ldots \altc \tau$ with 
$Y \not\equiv X$.
The corresponding part of the proof of 
Theorems~\ref{theorem-completeness-ACPet-1} 
and~\ref{theorem-completeness-ACPet-2} is likewise restricted to 
constants $\rec{X}{E}$ where $E$ does not contain equations 
$Y = \phi_1 \gc \tau \altc \ldots \altc \phi_n \gc \tau$ with 
$Y \not\equiv X$.
This is not because such an equation can be eliminated, but because it
can be replaced by $Y = \phi_1 \Lor \ldots \Lor \phi_n \gc \ep$.

The following is a corollary of Theorems~\ref{theorem-soundness-ACPet} 
and~\ref{theorem-completeness-ACPet-2}.
\begin{corollary}
\label{corollary-completeness-ACPet}
For all $t,t' \in \ProcTermr$, for all $\sigma \in \EvalMap$, 
$\deACPetrf \Ent \eval{\sigma}(t) = \eval{\sigma}(t')$ iff 
$\eval{\sigma}(t) \rbbisim \eval{\sigma}(t')$.
\end{corollary}

\section{Information-Flow Security}
\label{sect-info-flow}

In this section, it will be explained how \deACPet\ can be used for 
information-flow security analysis of the kind that is concerned with 
the leakage of confidential data.
However, first, a general idea is given of what information-flow 
security is about and what results have been produced by research on 
this subject.

Consider a program whose variables are partitioned into high-security 
variables and low-security variables. 
High-security variables are considered to contain confidential data and 
low-security variables are considered to contain non-confidential data. 
The information flow in the program is called secure if information 
derivable from data contained in the high-security variables cannot be 
inferred from data contained in the low-security variables. 
Secure information flow means that no confidential data is leaked.
A well-known program property that guarantees secure information flow is 
non-interference. 
In the case where the program is a deterministic sequential program, 
non-interference is the property that the data initially contained in 
high security variables has no effect on the data finally contained in 
low security variables. 

\sloppy
Theoretical work on information-flow security is already done since the
1970s (see e.g.~\cite{BP73a,Den76a,DD77a,Coh77a,Coh78a}).
A great part of the work done until now has been done in a 
programming-language setting. 
This work has among other things led to security-type systems for 
programming languages. 
The languages concerned vary from languages supporting sequential 
programming to languages supporting concurrent programming and from 
languages for programming transformational systems to languages for 
programming reactive systems 
(see e.g.~\cite{VIS96a,SV98a,BC02a,NCC06a,Boh09a}).

However, work on information-flow security has also been done in a 
process-algebra setting. 
In such a setting, the information flow in a process is generally called 
secure if information derivable from confidential actions cannot be 
inferred from non-confidential actions
(see e.g.~\cite{FG95a,RS99a,Bos04a,Low04a}). 
So, in a process-algebra setting, secure information flow usually means 
that no confidential action is revealed. 
Moreover, in such a setting, non-interference is the property that the 
confidential actions have no effect on the non-confidential actions. 
Recently, work done on information-flow security in a process-algebra 
setting occasionally deals with the data-oriented notion of secure 
information flow, but on such occasions program variables are always 
mimicked by processes (see e.g.~\cite{FRS05a,HY07a}). 
\deACPet\ obviates the need to mimic program variables.

In the rest of this section, the interest is in processes that are 
carried out by systems that have a state comprising a number of 
data-containing components whose content can be looked up and changed.
Moreover, the attention is focussed on processes, not necessarily 
arising from the execution of a program, in which (a)~confidential and 
non-confidential data contained in the state components of the system in
question are looked up and changed and (b)~an ongoing interaction with 
the environment of the system in question is maintained where data are 
communicated in either direction.
In the terminology of \deACPet, the state components are called flexible 
variables. 
From now on, processes of the kind described above are referred to as 
processes of the type of interest.
The processes that are carried out by many contemporary systems are 
covered by the processes of the type of interest. 

The point of view is taken that the information flow in a process of the 
type of interest is secure if information derivable from the 
confidential data contained in state components cannot be inferred from 
its interaction with the environment. 
A process property that guarantees secure information flow in this sense 
is the property that the confidential data contained in state components
has no effect on the interaction with the environment. 
This property, which will be made more precise below, is called the DNII 
(Data Non-Interference with Interactions) property.
For a process with this property, differences in the confidential data 
contained in state components cannot be observed in 
(a)~what remains of the process in the case where only the actions that 
are performed to interact with the environment are visible and 
(b)~consequently in the data communicated with the 
environment.

For each closed \deACPetr\ term $P$ of sort $\Proc$ that denotes a 
process of the type of interest, it is assumed that the following has 
been given:
\begin{itemize}
\item
a set $\LOW{P} \subseteq \ProgVar$ of \emph{low-security} flexible 
variables of $P$;
\item 
a set 
$\EXT{P} \subseteq 
\Act \union
\Union_{n \in \Natpos}
 \set{a(e_1,\dots,e_n) \where
      a \in \Act \Land e_1,\dots,e_n \in \DataVal}$ \linebreak[2]
of \emph{external} actions of $P$. 
\end{itemize}
For each closed \deACPetr\ term $P$ of sort $\Proc$ that denotes a 
process of the type of interest, we define the following sets:
\begin{ldispl}
\begin{aeqns}
\HIGH{P} & = &
\set{v \in \ProgVar^P \where v \notin \LOW{P}}\;, \\
\INT{P} & = &
\set{\alpha \in \AProcTerm^P \where \alpha \notin \EXT{P}}\;,
\end{aeqns}
\end{ldispl}%
where
\begin{ldispl}
\begin{aeqns}
\ProgVar^P & = &
\set{v \in \ProgVar \where v \mathrm{\;occurs\;in\;} P}\;, \\
\AProcTerm^P & = &
\set{a \in \Act \where a \mathrm{\;occurs\;in\;} P} \\
& & {} \union \,
\Union_{n \in \Natpos}
 \set{a(\sigma(e_1),\dots,\sigma(e_n)) \where
      \sigma \in \EvalMap \Land
      a(e_1,\dots,e_n) \mathrm{\;occurs\;in\;} P} \\
& & {} \union
\set{\ass{v}{\sigma(e)} \where
     \sigma \in \EvalMap \Land \ass{v}{e} \mathrm{\;occurs\;in\;} P}\;.
\end{aeqns}
\end{ldispl}%
$\HIGH{P}$ is called the set of \emph{high-security} flexible variables 
of $P$ and $\INT{P}$ is called the set of \emph{internal} actions of 
$P$. 

The flexible variables in $\LOW{P}$ are the flexible variables of $P$ 
that contain non-confidential data and the flexible variables in 
$\HIGH{P}$ are the flexible variables of $P$ that contain confidential 
data.
The actions in $\EXT{P}$ are the actions that are performed by $P$ to 
interact with the environment and the actions in $\INT{P}$ are the 
actions that are performed by $P$ to do something else than to interact 
with the environment.
The actions in $\INT{P}$ are considered to be invisible in the 
environment.
In earlier work based on a purely action-oriented notion of secure 
information flow, the actions in $\EXT{P}$ and $\INT{P}$ are called 
low-security actions and high-security actions, respectively, or 
something similar.

For each closed \deACPetr\ term $P$ of sort $\Proc$ that denotes a 
process of the type of interest, $P$ has the DNII property iff
\begin{ldispl}
\deACPetrf \Ent 
\abstr{\INT{P}}(\eval{\sigma}(P)) = \abstr{\INT{P}}(\eval{\sigma'}(P))
\end{ldispl}%
for all evaluation maps $\sigma$ and $\sigma'$ such that 
$\sigma(v) = \sigma'(v)$ for all $v \in \LOW{P}$.

This definition is justified by the fact, which follows from 
Corollary~\ref{corollary-completeness-ACPet}, that
\begin{ldispl}
\deACPetrf \Ent
\abstr{\INT{P}}(\eval{\sigma}(P)) = 
\abstr{\INT{P}}(\eval{\sigma'}(P)) \\
\hfill \mathrm{iff}\;\;
\abstr{\INT{P}}(\eval{\sigma}(P)) \rbbisim
\abstr{\INT{P}}(\eval{\sigma'}(P))\;. \hfill
\end{ldispl}%

The left-hand side and the right-hand side of the equation in the above
definition denote the processes that remain of the process denoted by 
$P$ in the case that the data values assigned to the flexible variables 
are initially as defined by $\sigma$ and $\sigma'$, respectively, and 
moreover all internal actions of $P$ are not visible.
The condition imposed on the evaluation maps $\sigma$ and $\sigma'$ 
tells us that the equation must always hold if, for each low-security 
flexible variable of $P$, the data values assigned to it according to 
$\sigma$ and $\sigma'$ are the same.
This corresponds to the intuitive idea mentioned above that, for a 
process with the DNII property, differences in the confidential data 
cannot be observed in what remains of the process in the case where only 
the actions that are performed to interact with the environment are 
visible.

Assume that $h,l \in \ProgVar$ and $a,b \in \Act$. 
Let $P$ be the closed \deACPetr\ term
\begin{ldispl}
(h = 0)       \gc \ass{l}{l + 1} \seqc a \altc 
\Lnot (h = 0) \gc a \seqc \ass{l}{l + 1} \altc b
\end{ldispl}%
of sort $\Proc$ with $\LOW{P} = \set{l}$ and $\EXT{P} = \set{a,b}$.
$P$ is a very simple example of a term of which it may not be 
immediately clear that it denotes a process that does not have the DNII 
property.
Notice that, by definition, $h \in \HIGH{P}$ and 
$\ass{l}{l + 1} \in \INT{P}$.
When $\ass{l}{l + 1}$ is performed, this cannot be observed in the 
externally observable process because $\ass{l}{l + 1}$ is an internal 
action.
This means that, irrespective of the value that is initially assigned to
$h$, the externally observable process performs either $a$ or $b$ and 
after that terminates successfully.
This is why the process denoted by $P$ may seem to have the DNII 
property.
However, at the point that the externally observable process has the 
option to perform $a$, it has also the option to perform $b$ in the case 
where the value initially assigned to $h$ is not $0$, while it does not 
have also the option to perform $b$ in the case where the value 
initially assigned to $h$ is $0$.
In other words, the externally observable process in the former case 
differs from the externally observable process in the latter case.
This means that, whether or not $0$ is initially assigned to $h$ can be 
inferred from the externally observable process.
Hence, the informal conclusion is that $P$ denotes a process that does 
not have the DNII property.
More formally, this conclusion follows from the definition of the DNII
property: we have 
\begin{ldispl}
\deACPetrf \Ent
 \abstr{\INT{P}}(\eval{\sigma}(P)) = \tau \seqc a \altc b
\end{ldispl}%
for all evaluation maps $\sigma$ such that $\sigma(h) = 0$
and we have
\begin{ldispl}
\deACPetrf \Ent
 \abstr{\INT{P}}(\eval{\sigma'}(P)) = a \altc b
\end{ldispl}%
for all evaluation maps $\sigma'$ such that $\sigma'(h) \neq 0$,
but we do not have
\begin{ldispl}
\deACPetrf \Ent \tau \seqc a \altc b = a \altc b\;.
\end{ldispl}%

In CSP$_\sigma$~\cite{CH09a}, presumably the only imperative process 
algebra that supports abstraction from actions that are considered not 
to be visible, the DNII property cannot be defined. 
The cause of this is that abstraction from actions that are 
considered not to be visible means in CSP$_\sigma$ that these actions
are simply removed.
Because of that processes such as $\abstr{\INT{P}}(\eval{\sigma}(P))$ 
and $\abstr{\INT{P}}(\eval{\sigma'}(P))$ from the example given above 
are equated.

Assume that $h,l \in \ProgVar$ and $a,b \in \Act$.
Let $Q$ be the closed \deACPetr\ term
\begin{ldispl}
(h = 0)       \gc \ass{l}{l + 1} \seqc a \altc 
\Lnot (h = 0) \gc \ass{l}{l - 1} \seqc a \altc b
\end{ldispl}%
of sort $\Proc$ with $\LOW{Q} = \set{l}$ and $\EXT{Q} = \set{a,b}$.
$Q$ is a very simple example of a term that denotes a process that has 
the DNII property.
This follows from the definition of the DNII property: we have
\begin{ldispl}
\deACPetrf \Ent
 \abstr{\INT{Q}}(\eval{\sigma}(Q)) = \tau \seqc a \altc b
\end{ldispl}%
for all evaluation maps $\sigma$ such that $\sigma(h) = 0$,
we have
\begin{ldispl}
\deACPetrf \Ent
 \abstr{\INT{Q}}(\eval{\sigma'}(Q)) = \tau \seqc a \altc b
\end{ldispl}%
for all evaluation maps $\sigma'$ such that $\sigma'(h) \neq 0$,
and we trivially have
\begin{ldispl}
\deACPetrf \Ent \tau \seqc a \altc b = \tau \seqc a \altc b\;.
\end{ldispl}%

The DNII property is only one of the process properties related to 
information flow security that can be defined and verified in 
\deACPetrf.
Insofar as information flow security of contemporary systems is 
concerned, it seems to be an essential property.
The DNII property is also one of the process properties related to 
information flow security that cannot be defined naturally in the 
process algebras used in earlier work on information flow security 
(cf.~\cite{FG95a,RS99a,Bos04a,Low04a}).
The problem with those process algebras is that state components must be
mimicked by processes in them.

The DNII property concerns the non-disclosure of confidential data, 
contained in state components of a system, through the possibly ongoing 
interaction of the system concerned with its environment.
To my knowledge, such a property has not been proposed in the literature 
on information-flow security before.
However, the DNII property is reminiscent of the combination of data 
non-interference and event non-interference as defined in~\cite{SC20a},
but a comparison is difficult to make because of rather different
semantical bases.

\section{Concluding Remarks}
\label{sect-conclusions}

I have introduced an \ACP-based imperative process algebra.
This process algebra distinguishes itself from imperative process 
algebras such as VPLA~\cite{HI93a}, IPAL~\cite{NP97a}, 
CSP$_\sigma$~\cite{CH09a}, AWN~\cite{Feh12a}, and the process algebra 
proposed in~\cite{BLSW20a} by the following three properties:
(1)~it supports abstraction from actions that are considered not to be 
visible;
(2)~a verification of the equivalence of two processes in its semantics 
is automatically valid in any semantics that is fully abstract with 
respect to some notion of observable behaviour;
(3)~it offers the possibility of equational verification of process
equivalence. 

Properties (1)--(3) have been achieved by the inclusion of the silent 
step constant $\tau$ and the abstraction operators $\abstr{I}$ in the 
constants and operators of the process algebra, the use of the rooted 
branching bisimulation equivalence relation $\rbbisim$ for the 
equivalence of processes in its semantics, and the provision of the 
equational axiomatization of~$\rbbisim$.

The axioms of the presented imperative process algebra are not complete 
with respect to the equivalence of processes in its semantics.
There is no straightforward way to rectify this.
However, two semi-completeness results that may be relevant to various 
applications of this imperative process algebra have been established.
One of those results is at least relevant to information-flow security 
analysis.
The finiteness and linearity restrictions on guarded recursive 
specifications are not needed for the uniqueness of solutions.
However, there would be no semi-completeness results without these 
restrictions. 

In this paper, I build on earlier work on ACP. 
The axioms of \ACPet\ have been taken from Section 5.3 of~\cite{BW90} 
and the axioms for the guarded command operator have been taken 
from~\cite{BB92c}.
The  evaluation operators have been inspired by~\cite{BM05a} and the 
data parameterized action operators have been inspired by~\cite{BM09d}.

\subsection*{Acknowledgement}

I thank two anonymous referees for carefully reading a preliminary 
version of this paper, for suggesting improvements of the presentation 
of the paper, and for pointing out two minor but annoying errors in it.

\appendix

\section*{Appendix: Alternative Bisimulation Semantics}
\label{sect-alt-semantics}

In this appendix, an alternative to the structural operational semantics 
of \mbox{\deACPetr}\ is presented and a definition of rooted branching 
bisimulation equivalence for \deACPetr\ based on this alternative 
structural operational semantics is given.
This appendix is strongly based on Section~4 of~\cite{BM19b}.

We write $\sCondTerm$ for the set of all terms $\phi \in \CondTerm$ for 
which $\nSat{\gD}{\phi \Liff \False}$.
As formulas of a first-order language with equality of $\gD$, the terms 
from $\sCondTerm$ are the formulas that are satisfiable in $\gD$.

The alternative structural operational semantics of \deACPetr\ consists 
of \begin{itemize}
\item 
a binary conditional transition relation \smash{$\step{\ell}$} on 
$\ProcTermr$ for each $\ell \in \sCondTerm \x \AProcTermt$;
\item 
a unary successful termination relation $\sterm{\phi}$ on $\ProcTermr$ 
for each $\phi \in \sCondTerm$.
\end{itemize}
We write \smash{$\astep{t}{\gact{\phi}{\alpha}}{t'}$} instead of 
\smash{$\tup{t,t'} \in {\step{\tup{\phi,\alpha}}}$} and 
$\isterm{t}{\phi}$ instead of $t \in {\sterm{\phi}}$.

The relations from this structural operational semantics describe what 
the processes denoted by terms from $\ProcTermr$ are capable of doing as 
follows:
\begin{itemize}
\item
$\astep{t}{\gact{\phi}{\alpha}}{t'}$: 
if condition $\phi$ holds for the process denoted by $t$, then this 
process has the potential to make a transition to the process denoted by 
$t'$ by performing action $\alpha$;
\item
$\isterm{t}{\phi}$: 
if condition $\phi$ holds for the process denoted by $t$, then this
process has the potential to terminate successfully.
\end{itemize}

The relations  from this structural operational semantics of \deACPetr\
are the smallest relations satisfying the rules given in 
Table~\ref{sos-deACPet-alt}.%
\begin{table}[!p]
\caption{Transition rules for \deACPet}
\label{sos-deACPet-alt}
\begin{ruletbl}
{} 
\Rule
{}
{\astep{\alpha}{\gact{\True}{\alpha}}{\ep}}
\quad\;
\Rule
{}
{\isterm{\ep}{\True}}
\quad\;
\Rule
{\isterm{x}{\phi}}
{\isterm{x \altc y}{\phi}}
\quad\;
\Rule
{\isterm{y}{\phi}}
{\isterm{x \altc y}{\phi}}
\quad\;
\Rule
{\astep{x}{\gact{\phi}{\alpha}}{x'}}
{\astep{x \altc y}{\gact{\phi}{\alpha}}{x'}}
\quad\;
\Rule
{\astep{y}{\gact{\phi}{\alpha}}{y'}}
{\astep{x \altc y}{\gact{\phi}{\alpha}}{y'}}
\\
\RuleC
{\isterm{x}{\phi},\; \isterm{y}{\psi}}
{\isterm{x \seqc y}{\phi \Land \psi}}
{\nSat{\gD}{\fol{\phi \Land \psi \Liff \False}}}
\quad\;
\RuleC
{\isterm{x}{\phi},\; \astep{y}{\gact{\psi}{\alpha}}{y'}}
{\astep{x \seqc y}{\gact{\phi \Land \psi}{\alpha}}{y'}}
{\nSat{\gD}{\fol{\phi \Land \psi \Liff \False}}}
\quad\;
\Rule
{\astep{x}{\gact{\phi}{\alpha}}{x'}}
{\astep{x \seqc y}{\gact{\phi}{\alpha}}{x' \seqc y}}
\\
\RuleC
{\isterm{x}{\phi},\; \isterm{y}{\psi}}
{\isterm{x \parc y}{\phi \Land \psi}}
{\nSat{\gD}{\fol{\phi \Land \psi \Liff \False}}}
\quad\;
\Rule
{\astep{x}{\gact{\phi}{\alpha}}{x'}}
{\astep{x \parc y}{\gact{\phi}{\alpha}}{x' \parc y}}
\quad\;
\Rule
{\astep{y}{\gact{\phi}{\alpha}}{y'}}
{\astep{x \parc y}{\gact{\phi}{\alpha}}{x \parc y'}}
\\
\RuleC
{\astep{x}{\gact{\phi}{a}}{x'},\; \astep{y}{\gact{\psi}{b}}{y'}}
{\astep{x \parc y}{\gact{\phi \Land \psi}{c}}{x' \parc y'}}
{\commf(a,b) = c,\; \nSat{\gD}{\fol{\phi \Land \psi \Liff \False}}}
\\
\RuleC
{\astep{x}{\gact{\phi}{a(e_1,\ldots,e_n)}}{x'},\; 
 \astep{y}{\gact{\psi}{b(e'_1,\ldots,e'_n)}}{y'}}
{\astep{x \parc y}
  {\gact{\phi \Land \psi \Land e_1 = e'_1 \Land \ldots \Land  e_n = e'_n}
  {c(e_1,\ldots,e_n)}}{x' \parc y'}}
{\begin{array}{@{}c@{}}
 \commf(a,b) = c, \\[.25ex] 
 \nSat{\gD}
  {\fol{\phi \Land \psi \Land e_1 = e'_1 \Land \ldots \Land  e_n = e'_n \Liff
        \False}}
 \end{array}
}
\\
\Rule
{\astep{x}{\gact{\phi}{\alpha}}{x'}}
{\astep{x \leftm y}{\gact{\phi}{\alpha}}{x' \parc y}}
\\
\RuleC
{\astep{x}{\gact{\phi}{a}}{x'},\; \astep{y}{\gact{\psi}{b}}{y'}}
{\astep{x \commm y}{\gact{\phi \Land \psi}{c}}{x' \parc y'}}
{\commf(a,b) = c,\; \nSat{\gD}{\fol{\phi \Land \psi \Liff \False}}}
\\
\RuleC
{\astep{x}{\gact{\phi}{a(e_1,\ldots,e_n)}}{x'},\; 
 \astep{y}{\gact{\psi}{b(e'_1,\ldots,e'_n)}}{y'}}
{\astep{x \commm y}
  {\gact{\phi \Land \psi \Land e_1 = e'_1 \Land \ldots \Land  e_n = e'_n}
  {c(e_1,\ldots,e_n)}}{x' \parc y'}}
{\begin{array}{@{}c@{}}
 \commf(a,b) = c, \\[.25ex]
 \nSat{\gD}
  {\fol{\phi \Land \psi \Land e_1 = e'_1 \Land \ldots \Land  e_n = e'_n \Liff
        \False}}
 \end{array}
}
\\
\Rule
{\isterm{x}{\phi}}
{\isterm{\encap{H}(x)}{\phi}}
\quad\;
\RuleC
{\astep{x}{\gact{\phi}{\alpha}}{x'}}
{\astep{\encap{H}(x)}{\gact{\phi}{\alpha}}{\encap{H}(x')}}
{\alpha \notin H}
\\
\Rule
{\isterm{x}{\phi}}
{\isterm{\abstr{I}(x)}{\phi}}
\quad\;
\RuleC
{\astep{x}{\gact{\phi}{\alpha}}{x'}}
{\astep{\abstr{I}(x)}{\gact{\phi}{\alpha}}{\abstr{I}(x')}}
{\alpha \notin I}
\quad\;
\RuleC
{\astep{x}{\gact{\phi}{\alpha}}{x'}}
{\astep{\abstr{I}(x)}{\gact{\phi}{\tau}}{\abstr{I}(x')}}
{\alpha \in I}
\\
\RuleC
{\isterm{x}{\phi}}
{\isterm{\psi \gc x}{\phi \Land \psi}}
{\nSat{\gD}{\fol{\phi \Land \psi \Liff \False}}}
\quad\;
\RuleC
{\astep{x}{\gact{\phi}{\alpha}}{x'}}
{\astep{\psi \gc x}{\gact{\phi \Land \psi}{\alpha}}{x'}}
{\nSat{\gD}{\fol{\phi \Land \psi \Liff \False}}}
\\
\RuleC
{\isterm{x}{\phi}}
{\isterm{\eval{\sigma}(x)}{\sigma(\phi)}}
{\nSat{\gD}{\fol{\sigma(\phi) \Liff \False}}}
\quad\;
\RuleC
{\astep{x}{\gact{\phi}{\tau}}{x'}}
{\astep{\eval{\sigma}(x)}{\gact{\sigma(\phi)}{\tau}}{\eval{\sigma}(x')}}
{\nSat{\gD}{\fol{\sigma(\phi) \Liff \False}}}
\\
\RuleC
{\astep{x}{\gact{\phi}{a}}{x'}}
{\astep{\eval{\sigma}(x)}{\gact{\sigma(\phi)}{a}}{\eval{\sigma}(x')}}
{\nSat{\gD}{\fol{\sigma(\phi) \Liff \False}}}
\quad\;
\RuleC
{\astep{x}{\gact{\phi}{a(e_1,\ldots,e_n)}}{x'}}
{\astep{\eval{\sigma}(x)}
       {\gact{\sigma(\phi)}{a(\sigma(e_1),\ldots,\sigma(e_n))}}
       {\eval{\sigma}(x')}}
{\nSat{\gD}{\fol{\sigma(\phi) \Liff \False}}}
\\
\RuleC
{\astep{x}{\gact{\phi}{\ass{v}{e}}}{x'}}
{\astep{\eval{\sigma}(x)}{\gact{\sigma(\phi)}{\ass{v}{\sigma(e)}}}
       {\eval{\sigma\mapupd{\sigma(e)}{v}}(x')}}
{\nSat{\gD}{\fol{\sigma(\phi) \Liff \False}}}
\\
\RuleC
{\isterm{\rec{t}{E}}{\phi}}
{\isterm{\rec{X}{E}}{\phi}}
{X \!\!=\! t \,\in\, E}
\qquad
\RuleC
{\astep{\rec{t}{E}}{\gact{\phi}{\alpha}}{x'}}
{\astep{\rec{X}{E}}{\gact{\phi}{\alpha}}{x'}}
{X \!\!=\! t \,\in\, E}
\vspace*{1ex}
\end{ruletbl}
\end{table}
In this table, 
$\alpha$ stands for an arbitrary action from $\AProcTermt$,\,
$\phi$ and $\psi$ stand for arbitrary terms from $\sCondTerm$,\,
$a,b$, and $c$ stand for arbitrary basic actions from $\Act$,
\linebreak[2]
$e,e_1,e_2,\ldots$ and $e'_1,e'_2,\ldots$ stand for arbitrary terms 
from $\DataTerm$,\,
$H$ stands for an arbitrary subset of $\AProcTerm$ or the set 
$\AProcTermt$,\, 
$I$ stands for an arbitrary subset of~$\AProcTerm$,\,
$\sigma$ stands for an arbitrary evaluation map from $\EvalMap$,\,
$v$ stands for an arbitrary flexible variable from $\ProgVar$,\,
$X$ stands for an arbitrary variable from $\cX$,\, 
$t$ stands for an arbitrary \deACPet\ term of sort $\Proc$, and
$E$ stands for an arbitrary guarded linear recursive specification over 
\deACPet.

The alternative structural operational semantics is such that the 
structural operational semantics presented in 
Section~\ref{sect-semantics} can be obtained by replacing each 
transition \smash{$\astep{t}{\gact{\phi}{\alpha}}{t'}$} by a transition 
\smash{$\astep{t}{\gact{\sigma}{\alpha}}{t'}$} for each 
$\sigma \in \EvalMap$ for which $\Sat{\gD}{\sigma(\phi)}$, and 
likewise each \smash{$\isterm{t}{\phi}$}.

Two processes are considered equal if they can simulate each other 
insofar as their observable potentials to make transitions and to 
terminate successfully are concerned.
In the case of the alternative structural operational semantics, there 
are two issues that together complicate matters:
\begin{itemize}
\item
simply relating a single transition of one of the processes to a single 
transition of the other process does not work because a transition of 
one process may be simulated by a set of transitions of another process;
\item
simply ignoring all transitions in which the unobservable action $\tau$ 
is performed does not work because the observable potentials to make 
transitions and to terminate successfully may change by such 
transitions.
\end{itemize}
The first issue is illustrated by the processes denoted by
$\phi \Lor \psi \gc a$ and $\phi \gc a \altc \psi \gc a$:
the only transition of the former process is simulated by the two 
transitions of the latter process.
The second issue is illustrated by the processes denoted by
$a \altc \tau \seqc b$ and $a \altc b$:
by making the transition in which the unobservable action $\tau$ is 
performed, the former process loses the potential to make the 
transition in which the observable action $a$ is performed before 
anything has been observed, whereas this potential is a potential of the 
latter process so long as nothing has been observed. 

The first issue alone can be dealt with by means of the notion of 
splitting bisimulation equivalence introduced in~\cite{BM05a} and the 
second issue alone can be dealt with by means of the notion of branching 
bisimulation equivalence introduced in~\cite{GW96a} adapted to the 
conditionality of transitions in which the unobservable action $\tau$ is 
performed.
In order to deal with both issues, the two notions are combined.

We write \smash{$\astep{t}{(\gact{\phi}{\alpha})}{t'}$}, 
where $\phi \in \sCondTerm$ and $\alpha \in \AProcTermt$, 
for \smash{$\astep{t}{\gact{\phi}{\alpha}}{t'}$} or 
$\alpha = \tau$, $t = t'$, and $\Sat{\gD}{\phi \Liff \True}$.

The notation $\LOR \Phi$, where $\Phi = \set{\phi_1,\ldots,\phi_n}$ and 
$\phi_1, \ldots, \phi_n$ are \deACPet\ terms of sort $\Cond$, is used 
for the \deACPet\ term $\phi_1 \Lor \ldots \Lor \phi_n$. 

An \emph{ab-bisimulation} is a binary relation $R$ on $\ProcTermr$ such 
that, for all terms $t_1,t_2 \in \ProcTermr$ with $(t_1,t_2) \in R$, the 
following transfer conditions hold:
\begin{itemize}
\item
if $\astep{t_1}{\gact{\phi}{\alpha}}{t_1'}$, then 
there exists a finite set $\Psi \subseteq \sCondTerm$ such that 
\mbox{$\Sat{\gD}{\phi \Limpl \LOR \Psi}$} and, 
for all $\psi \in \Psi$, there exists an $\alpha' \in [\alpha]$ and, 
for some $n \in \Nat$, there exist
$t_2^0,\ldots,t_2^n,t_2' \in \ProcTermr$ and 
$\psi^1,\ldots,\psi^n,\psi' \in \sCondTerm$ such that 
$\Sat{\gD}{\psi \Liff \psi' \Land \psi^1 \Land \ldots \Land \psi^n}$,
$t_2^0 \equiv t_2$, 
for all $i \in \Nat$ with $i < n$,
$\astep{t_2^i}{\gact{\psi^{i+1}}{\tau}}{t_2^{i+1}}$ and
$(t_1,t_2^{i+1}) \in R$,
$\astep{t_2^n}{(\gact{\psi'}{\alpha'})}{t_2'}$, 
and $(t_1',t_2') \in R$;
\item
if $\astep{t_2}{\gact{\phi}{\alpha}}{t_2'}$, then 
there exists a finite set $\Psi \subseteq \sCondTerm$ such that 
\mbox{$\Sat{\gD}{\phi \Limpl \LOR \Psi}$} and, 
for all $\psi \in \Psi$, there exists an $\alpha' \in [\alpha]$ and, 
for some $n \in \Nat$, there exist
$t_1^0,\ldots,t_1^n,t_1' \in \ProcTermr$ and 
$\psi^1,\ldots,\psi^n,\psi' \in \sCondTerm$ such that 
$\Sat{\gD}{\psi \Liff \psi' \Land \psi^1 \Land \ldots \Land \psi^n}$,
$t_1^0 \equiv t_1$, 
for all $i \in \Nat$ with $i < n$,
$\astep{t_1^i}{\gact{\psi^{i+1}}{\tau}}{t_1^{i+1}}$ and
$(t_1^{i+1},t_2) \in R$,
$\astep{t_1^n}{(\gact{\psi'}{\alpha'})}{t_1'}$, 
and $(t_1',t_2') \in R$;
\item
if $\isterm{t_1}{\phi}$, then 
there exists a finite set $\Psi \subseteq \sCondTerm$ such that 
\mbox{$\Sat{\gD}{\phi \Limpl \LOR \Psi}$} and, 
for all $\psi \in \Psi$, 
for some $n \in \Nat$, there exist
$t_2^0,\ldots,t_2^n \in \ProcTermr$ and 
$\psi^1,\ldots,\psi^n,\psi' \in \sCondTerm$ such that 
$\Sat{\gD}{\psi \Liff \psi' \Land \psi^1 \Land \ldots \Land \psi^n}$,
$t_2^0 \equiv t_2$, 
for all $i \in \Nat$ with $i < n$,
$\astep{t_2^i}{\gact{\psi^{i+1}}{\tau}}{t_2^{i+1}}$ and
$(t_1,t_2^{i+1}) \in R$, and $\isterm{t_2^n}{\psi'}$;
\item
if $\isterm{t_2}{\phi}$, then 
there exists a finite set $\Psi \subseteq \sCondTerm$ such that 
\mbox{$\Sat{\gD}{\phi \Limpl \LOR \Psi}$} and, 
for all $\psi \in \Psi$, 
for some $n \in \Nat$, there exist
$t_1^0,\ldots,t_1^n \in \ProcTermr$ and 
$\psi^1,\ldots,\psi^n,\psi' \in \sCondTerm$ such that 
$\Sat{\gD}{\psi \Liff \psi' \Land \psi^1 \Land \ldots \Land \psi^n}$,
$t_1^0 \equiv t_1$, 
for all $i \in \Nat$ with $i < n$,
$\astep{t_1^i}{\gact{\psi^{i+1}}{\tau}}{t_1^{i+1}}$ and
$(t_1^{i+1},t_2) \in R$, and $\isterm{t_1^n}{\psi'}$.
\end{itemize}
If $R$ is an ab-bisimulation, then a pair $(t_1,t_2)$  is said to
satisfy the root condition in $R$ if the following conditions hold:
\begin{itemize}
\item
if $\astep{t_1}{\gact{\phi}{\alpha}}{t_1'}$, then 
there exists a finite set $\Psi \subseteq \sCondTerm$ such that 
\mbox{$\Sat{\gD}{\phi \Limpl \LOR \Psi}$} and, 
for all $\psi \in \Psi$, there exist an $\alpha' \in [\alpha]$ and  
a $t_2' \in \ProcTermr$ such that 
\smash{$\astep{t_2}{\gact{\psi}{\alpha'}}{t_2'}$} and 
$(t_1',t_2') \in R$;
\item
if $\astep{t_2}{\gact{\phi}{\alpha}}{t_2'}$, then 
there exists a finite set $\Psi \subseteq \sCondTerm$ such that 
\mbox{$\Sat{\gD}{\phi \Limpl \LOR \Psi}$} and, 
for all $\psi \in \Psi$, there exist an $\alpha' \in [\alpha]$ and  
a $t_1' \in \ProcTermr$ such that 
\smash{$\astep{t_1}{\gact{\psi}{\alpha'}}{t_1'}$} and 
$(t_1',t_2') \in R$;
\item
if $\isterm{t_1}{\phi}$, then 
there exists a finite set $\Psi \subseteq \sCondTerm$ such that 
$\Sat{\gD}{\phi \Limpl \LOR \Psi}$ and, 
for all $\psi \in \Psi$, $\isterm{t_2}{\psi}$;
\item
if $\isterm{t_2}{\phi}$, then 
there exists a finite set $\Psi \subseteq \sCondTerm$ such that 
$\Sat{\gD}{\phi \Limpl \LOR \Psi}$ and, 
for all $\psi \in \Psi$, $\isterm{t_1}{\psi}$.
\end{itemize}
Two terms $t_1,t_2 \in \ProcTermr$ are \emph{rooted ab-bisimulation 
equivalent}, written \mbox{$t_1 \rabbisim t_2$}, if there exists an 
ab-bisimulation $R$ such that $(t_1,t_2) \in R$ and $(t_1,t_2)$ 
satisfies the root condition in $R$.

In the absence of the constant $\tau$, rooted ab-bisimulation 
equivalence is essentially the same as splitting bisimulation 
equivalence as defined in~\cite{BM05a}.
In the absence of all terms of sort $\Cond$ other than the constants 
$\True$ and $\False$, rooted ab-bisimulation equivalence is essentially 
the same as rooted branching bisimulation equivalence as defined 
in~\cite{GW96a}.

I conjecture that, for all terms $t_1,t_2 \in \ProcTermr$, 
$t_1 \rbbisim t_2$ iff $t_1 \rabbisim t_2$.

\bibliographystyle{plain}
\bibliography{PA}

\end{document}